\documentclass[12pt,preprint]{aastex}

\def\F0{F_{\rm 0}}
\def\t0{t_{\rm 0}}
\def\NE{N_{\rm E}}

\newcommand{\ltsima} {$\; \buildrel < \over \sim \;$}
\newcommand{\gtsima} {$\; \buildrel > \over \sim \;$}
\newcommand{\lta} {\lower.5ex\hbox{\ltsima}}
\newcommand{\gta} {\lower.5ex\hbox{\gtsima}}

\slugcomment{Date: \today}
\begin{document}

\title{Quasi-blackbody component and radiative efficiency
of the prompt emission of  gamma-ray bursts}
 
\author{Felix Ryde}
\affil{Department of Physics, Royal Institute of Technology,
AlbaNova, SE-106 91 Stockholm, Sweden}

\author{Asaf Pe'er}
\affil{Space Telescope Science Institute, 3700 San Martin Drive, Baltimore, MD 21218}
\affil{Riccardo Giacconi fellow}

\begin{abstract}

We perform time-resolved spectroscopy on the prompt emission in gamma-ray bursts (GRBs) and identify a thermal, photospheric component peaking at a temperature of a few hundreds keV. This peak does not necessarily coincide with the broad band (keV-GeV) power peak.  We show that this thermal component exhibits a characteristic temporal behavior. We study a sample of 56 long bursts, all strong enough to allow time-resolved spectroscopy.  We analyze the evolution of both the temperature and flux of the thermal component in 49 individual time-resolved pulses, for which the temporal coverage is sufficient,  and find  that  the temperature is nearly constant during the first few seconds, after which it decays as a power law with a sample-averaged index of $-0.68$. The thermal flux first rises with an averaged power-law index of 0.63 after which it decays with an averaged index of $-2$. The  break times are the same to within errors. We find that the ratio of the observed to the emergent thermal flux typically exhibits a monotoneous  power-law increase during the entire pulse as well as during complex bursts.  Thermal photons carry a significant fraction ($\sim 30 \%$ to more than $50\%$) of the prompt emission energy (in the observed 25-1900 keV energy band), thereby significantly contributing to  the high radiative efficiency.  Finally, we show here that the thermal emission can be used to study the properties of the photosphere, hence the physical parameters of the GRB fireball. 
\end{abstract}


\section{Introduction}

Although being studied for several decades now, observations of the prompt emission phase of gamma-ray bursts (GRBs) have not yet been able to lead us to a clear understanding of either the underlying mechanism of the emission, nor its origin. The prompt emission spectrum is commonly modeled as a smoothly broken-power law \citep[which has become known as the Band function; ][]{Band93, Preece98a, Preece00, Barraud03, Kaneko06, Kaneko08, Abdo09, gonz09}.  While being phenomenological in nature, in many cases this interpretation was found to be consistent with the predictions of optically-thin
synchrotron emission from a power-law distribution of  energetic electrons \citep{Tavani96, Cohen97, Schaefer98, Frontera00, Wang09}. 
A peak in the energy spectrum then naturally arises, connected to the low-energy cut in the distribution of electrons. Indeed, it is widely accepted that the non-thermal spectra arise from the prompt
dissipation of a substantial fraction of the bulk kinetic energy of a relativistic outflow, originating from a central compact object
\citep[for review, see, e.g.,][]{Mes06}. The dissipated energy is converted into acceleration of 
electrons, which produce high-energy photons by synchrotron emission and inverse Compton (IC) scattering.

In spite of its many successes, evidence have accumulated in recent years  for low-energy spectral slopes  that are steeper than allowed by the optically-thin synchrotron or synchrotron self-Compton (SSC) model predictions \citep{Crider97, Preece98b, Preece02, ghirlanda, gonz09}. Moreover, the fitting is often made to the \emph{ time-integrated} spectrum.  Analysis of \emph{time-resolved} spectra done by \citet{CLP98} and  \citet{ghirlanda}, showed that neither the synchrotron nor the synchrotron self-Compton models can explain the instantaneous GRB spectra and their evolution. Motivated by these findings, other origins of the emission has been suggested such as small-pitch-angle synchrotron emission (Epstein \& Petrosian 1973, Medvedev 2006) and inverse Compton scattering of a self-aborbed seed emission. However, as  \citet{BaringB} have pointed out these emission mechanisms rely on almost purely non-thermal distributions of electrons  to be able to fit the observed spectra (see also \citet{SP04}). This is not a realistic assumption since diffusive shock acceleration typically leads to  a strong contribution of a thermal population. This was recently shown in the works by, for instance, \citet{spit08, gi09}. In addition, internal shocks are assumed to dissipate the bulk kinetic energy, in order for it to be emitted by synchrotron emission. This process is however too inefficient to explain the observed energy release (\citet{K97,L99,N08})  and moreover  the predicted correlation between peak energy and luminosity, that such a scenario gives rise to, contradicts what is observed (Ramirez-Ruiz \& Lloyd-Ronning 2002).

The difficulties that the purely non-thermal  emission models face are readily alleviated by introducing an additional optically-thick thermal (blackbody) component that may contribute to the observed spectrum \citep{MR00, MRRZ02, DM02, RM05}.  Two different emission regions contribute in forming the spectrum as well as its temporal evolution;  a narrow thermal component and a broad, non-thermal component are thus combined.   In this interpretation, the prominent spectral break (peak) in the sub-MeV range is identified with the peak  of the thermal emission, instead of being related to the  radiating electrons at the low-energy cut in their energy distribution\footnote{We further discuss the difference and relation between our interpretation and the traditional Band model fits in S \ref{sec:nonthermal}}. This has significant  advantages since it can accommodate (i) the existence of hard subpeak spectral slopes, (ii) temporal variations of the spectral shape, (iii) the observed spectral correlation between peak energy and luminosity within a burst (Golenetski et al. 1983), (iv) the high radiative efficiency required, and (v) the existence of complex spectral shapes,  such as Figure 1 and Figs. 5 and 6 in Ryde et al. (2006). In addition, the thermal peak-energy is weakly dependent on the luminosity  ($T \propto \Gamma L^{1/4}$) while the synchrotron peak-energy  is much more sensitive to uncertain values of the magnetic field and of the energy in the electrons ($E_{\rm p}  \propto  \Gamma\gamma_{\rm min}^2 B')$ and therefore expected to have a large dispersion. 

Indeed, from a theoretical point-of-view, thermal radiation originating from the base of the relativistic flow, where the densities are high enough to provide thermal equilibrium is an inevitable ingredient. This results from the fact that the fraction of  explosion energy that is being converted to kinetic energy (that is later dissipated to produce the non-thermal spectrum) is necessarily less than $100 \%$. Therefore, some fraction of the explosion energy is released in the form of radiation that thermalizes before escaping the flow at  the photosphere.  In addition, photons produced by any dissipation of the kinetic energy that occurs deep enough below the photosphere thermalize before escaping. The fraction of thermal emission that  dissipative outflow gives rise to depends mainly on (i) the relation between the radius at which the flow saturates (thermal and kinetic energy are equal) and the photospheric radius, (ii) the efficiency of the non-thermal energy dissipation, and  (iii) at what optical depth most of the  dissipation occurs. In other words the existence of a thermal component, which  is a natural outcome of relativistic outflow models, does not contradict the existence of a non-thermal emission, but adds to it.  

The interpretation of the prompt emission spectrum as being composed of a thermal component in addition to the non-thermal one, was put forward by \citet{Ryde04}.  In this work,  analysis of the time-resolved spectra of nine bright GRBs which  were characterized by hard low-energy  spectral slope, showed that a dominant thermal component could explain the observed spectra. The change in spectral shape over time, for instance by the low-energy spectral slope getting softer, was shown to be due to varying ratio of thermal to non-thermal flux and a change in the non-thermal spectral slope. This provides a natural explanation to the fact that  the observed spectra are hardest during the initial phases of the pulse. Indeed, \citet{ghirlanda} already suggested  that time-resolved spectra  during the initial  phases of some of the GRB pulses should be interpreted as pure  blackbody emission. 

In \citet{Ryde04} it was found that the observed\footnote{In the following, all quantities are in the observer frame unless otherwise stated. E.g. quantities in the comoving frame are primed.}  temperature of the thermal component is approximately constant during the first few  seconds, after which it decays as a power law in time $T \equiv T(t) \propto t^{a_T}$, with power-law index $a_T$  ranging approximately from $-0.6$  to  $-1.1$. It was later suggested \citep{Ryde05} that the thermal emission component can be identified in many other  bursts, in which it is not necessarily dominant over the non-thermal component.

In this paper, we put forward the idea suggested by \citet{Ryde05}, about the ubiquitousness of the thermal emission component in the prompt emission phase of GRBs. By using a large sample of 56 GRBs (see \S \ref{sec:data} for selection criteria) for which time resolved spectroscopy can be performed, we search  for the existence of a thermal emission component and study its  temporal behavior. In all the cases studied we were able to identify a thermal emission component.   We extend here previous analyses of the blackbody (photospheric) emission. While former works were focused only on the temperature evolution,  our present analysis  includes the evolution of both the temperature and thermal energy  flux, as well as the normalization of the blackbody fit, using a much larger sample than previously. This normalization corresponds to the ratio between the observed energy flux  and the emergent flux from the photosphere, $\sigma T^4$, where $\sigma$ is  Stefan-Boltzmann's constant.  As we show below, it can be interpreted as the emitting surface, that is, the surface of the photosphere. We present the analysis method in \S 2 and summarize our results, in particular the spectral evolution  of the blackbody component in \S 3.  In \S4, we discuss an interpretation of our results and elude on possible scenarios that could underly  the observed properties. We summarize and conclude in \S 5.

\section{Analysis Method}
\label{sec:data}
To properly analyze and determine the actual shape of GRB spectra one ideally needs burst detections with (i) high flux levels (ii) high time-resolution over individual pulses (iii) broad spectral coverage. Unfortunately, as of yet, all of these criteria are hard to meet at the same time, and the elucidation of the radiation process has not yet given a convincing answer. The best data for these investigations is still the BATSE sample from the {\it Compton Gamma-Ray Observatory}, which allows the analysis of spectra on short time scales over a spectral range of two order of magnitude above $\sim 25$ keV.  Here, we focus on the study of individual pulse structures within a GRB light curves, since these are the basic constituents of GRB light curve \citep{Band93}.  We select  our sample from bursts in the Kaneko spectral catalogue \citep{Kaneko06}. In order for a burst to be selected into  our sample it is required to have a single or a couple of distinct pulses, which do not overlap each other significantly. Furthermore, we use a  signal-to-noise ratio of more than 25 for the spectral analysis, which consequently determines the number and width of the individual time bins that are analyzed.
The pulses are then required to have enough time bins for our temporal analysis to be meaningful, implying more than six time bins over the analyzed duration. We also focus here only on the category of long bursts, longer than 2  seconds \citep{kouv, horvath}.  Our  sample then consists of 49 pulses in 48 bursts (2 pulses are analyzed in BATSE trigger 2083).  In addition to the complete sample of individual pulses, in order to demonstrate the robustness of our analysis method, we added to our sample eight bright bursts which have complex light curves with several heavily overlapping pulses. These are presented to illustrate the temporal behavior of the thermal component over the whole burst rather than only over a single pulse.  The only requirement for these additional bursts is their brightness. The sample here is used for demonstration purposes, and a full sample of such bursts will be published elsewhere. In total we thus present the analysis of  56 long bursts, for which good time resolved spectra exist. The full sample is presented in Table 1.

In the investigation below we follow the analysis by Ryde (2004, 2005) in which the background-subtracted photon
spectra, $ \NE(E,t)$, are fitted by a Planck function combined with a power-law:
\begin{equation}
\NE (E,t) = A(t) \,\, \frac{ E^2}{exp[E/kT(t)]-1}  +B(t) \,\, E^{s} 
= A(t) \,[kT]^2\,
\frac{x^2}{(e^x - 1)}+B(t) \,\,E^{s} , \label{eq:BBph}
\label{eq:BB}
\end{equation}
\noindent where $x \equiv E/k T$, $kT$ is the color temperature of the blackbody, and $k$ is Boltzmann's constant. 
The fitted parameters are thus the blackbody normalization $A(t)$, temperature $T(t)$, power-law index, $s(t)$ and its normalization, $B(t)$. The quality of the fit is given by a reduced $\chi_{\nu}^2$-value. In this paper, we restrict our study to the thermal component, since it is the best constrained component and has a direct  physical meaning (see \S 4 below). The non-thermal component is approximated by a single power-law over the analyzed energy range (25 -1900 keV). This assumption is, of course,  too simplified. Nontheless it can be justified over the limited BATSE energy band. We further discuss this assumption in 
\S \ref{sec:nonthermal} below; however a full, comprehensive study of the non-thermal component will be presented in a future paper.

The spectrum of the blackbody contribution to the observed flux  is modeled with the first term in equation (\ref{eq:BB}). At every instance (time bin) a temperature, $T(t)$, and a normalization, $A(t)$, are determined.  The energy flux, $F_{\rm BB}(t)= \int E N_{\rm E} \, dE$, is given by an integration over all energies of this term after multiplying it by $E(t)$. This integration yields 
$F_{\rm BB}(t) = A(t) \, [kT]^4 \, \pi^4/15$. We next analyze the temporal evolution of the parameters $T(t)$ and $F_{\rm BB}(t)$. We find that in 49 individual pulses it is possible to model the temporal evolution of the temperature with a broken power law,   $T\propto t^{a_T}$ before the break and    $T \propto t^{b_T}$  after the break. We thus find it possible to extend the analysis by Ryde 2004 to a larger sample.  As we show in \S 3, the temporal evolution of the flux can also be described by a broken power law, $F_{\rm BB}  \propto t^{a_F}$ before the break and   $F_{\rm BB}  \propto t^{b_F}$  after the break. Below we use the smoothly broken power law model derived in \citet{Ryde04} and described by their equation (6) for the temporal evolution for both these parameters. The other bursts in our sample are complex and have heavily overlapping pulses and therefore an analysis of individual pulses is not possible.  As described above for these bursts we make a similar analysis over the full burst duration.

The data we analyze was taken by the Large Area Detectors (LADs) on BATSE.  The data type we mainly use are the high-energy resolution burst (HERB) data \citep{Fish89}. These data have 128 energy channels over the energy range from   10 keV to a few MeV. The time resolution is multiples of 64 ms.  The background  estimates were made with the HER data, with a time resolution  of 16--500~s. The spectral analysis was performed with the RMFIT package, version  1.0b1 (Malozzi et~al. 2000) developed by the BATSE team in Huntsville, Alabama.  The  photon spectrum, $\NE(E)$ was determined using a forward-folding
technique; the spectral model was folded through the detector response matrix  and was then fitted by minimizing the
$\chi ^2$, using the non-linear Levenberg-Marquardt algorithm, between the model count spectrum and the observed count spectrum. This then gives the best-fit spectral parameters and the normalization. From these parameters the derived  quantities, such as the energy spectrum and the energy flux were calculated. In the following the bursts will be 
 identified by their BATSE trigger numbers according to the BATSE catalogue \citep{batse}, see Table 1.

\tabletypesize{\tiny}

\begin{deluxetable}{lccccccccl}
 \tablecolumns{10}
 \tablewidth{0pc}
 \tablecaption{Temporal evolution of the thermal component in a sample of 56 bursts. Listed for the 49 studied pulses within these bursts are the power law indices $a_T$ ($a_F$) and  $b_T$  ($b_F$)  of the temporal evolution of the temperature (flux) before and after the break time $t_0$, as well as the power law index $r$ of the temporal evolution of ${\cal{R}}$. See the text for details.}
\tablehead{ \colhead{Burst }  & \colhead{Trigger} & \colhead{$a_{T}$} & \colhead{$b_{T}$}
&\colhead{$t_{0, T}$} & \colhead{$a_{\rm F}$} & \colhead{$b_{\rm F}$} & \colhead{$t_{0, F}$} & \colhead{$r$} & \colhead{$\rm{}$}\\
\colhead{} &\colhead{} & \colhead{ } & \colhead{ } &\colhead{[s] } &
\colhead{ } & \colhead{} & \colhead{$[s]$} & \colhead{$$} & \colhead{${\rm Comments}$}
}

\startdata
910627	&	451	&	$	-0.2	\pm	0.67	$ & $	-0.64	\pm	-0.32	$ & $	1.07	\pm	1.3	$ & $	0.55	\pm	0.36	$ & $	-2.67	\pm	0.77	$ & $	1.73	\pm	0.32	$ & $	0.78	\pm	0.25	$	&	a, pulse 2	\\
910807	&	647	&	$	-0.08	\pm	0.05	$ & $	-0.77	\pm	0.13	$ & $	2.95	\pm	0.42	$ & $	0.5	\pm	0.12	$ & $	-2.24	\pm	0.33	$ & $	3.3	\pm	0.2	$ & $	0.52	\pm	0.05	$	&		\\
910814	&	678	&	$	0.001	\pm	0.13	$ & $	-0.43	\pm	0.043	$ & $	1.7	\pm	0.6	$ & $	0.29	\pm	0.28	$ & $	-1.43	\pm	0.13	$ & $	1.67	\pm	0.31	$ & $	0.02	\pm	0.03	$	&		\\
910927	&	829	&	$				$ & $				$ & $				$ & $				$ & $				$ & $				$ & $	0.63	\pm	0.05	$	&	Complex; Fig. \ref{fig:R1}	\\
911016	&	907	&	$	-0.18	\pm	0.05	$ & $	-0.58	\pm	0.02	$ & $	1.77	\pm	0.19	$ & $	0.27	\pm	0.12	$ & $	-1.26	\pm	0.06	$ & $	1.82	\pm	0.17	$ & $	0.54	\pm	0.03	$	&		\\
911031	&	973	&	$	-0.19	\pm	0.09	$ & $	-0.42	\pm	0.06	$ & $	2.7	\pm	1.2	$ & $	0.62	\pm	0.12	$ & $	-1.66	\pm	0.18	$ & $	2.85	\pm	0.2	$ & $	0.54	\pm	0.11	$	&	a, pulse 1	\\
911118	&	1085	&	$				$ & $				$ & $				$ & $				$ & $				$ & $				$ & $	0.82	\pm	0.03	$	&	Complex; Fig. \ref{fig:R1}	\\
920525	&	1625	&	$	0.4	\pm	0.2	$ & $	-0.85	\pm	0.1	$ & $	1.07	\pm	0.12	$ & $	2.6	\pm	0.6	$ & $	-3.5	\pm	0.2	$ & $	1.27	\pm	0.06	$ & $	0.39	\pm	0.12	$	&		\\
920622	&	1663	&	$				$ & $	          	$ & $				$ & $				$ & $				$ & $				$ & $				$	&	Complex; Fig. \ref{fig:R1}		\\
920718	&	1709	&	$	0	\pm	0.5	$ & $	-1.2	\pm	1.1	$ & $	1.04	\pm	0.9	$ & $	0.48	\pm	0.9	$ & $	-2.2	\pm	0.8	$ & $	0.69	\pm	0.3	$ & $	0.38	\pm	0.09	$	&	pulse 2	\\
920830	&	1883	&	$				$ & $	-0.43	\pm	0.06	$ & $				$ & $	0	\pm	0.1	$ & $	-2	\pm	0.2	$ & $	1.45	\pm	0.13	$ & $	0.71	\pm	0.04	$	&	b	\\
921003	&	1974	&	$	-0.05	\pm	0.24	$ & $	-0.32	\pm	0.03	$ & $	1.02	\pm	0.62	$ & $	1.2	\pm	0.2	$ & $	-0.8	\pm	0.23	$ & $	1.33	\pm	0.11	$ & $	0.61	\pm	0.05	$	&		\\
921123	&	2067	&	$ $ & $	 $ & $		$ & $		$ & $		$ & $	$ & $		$	& Complex;  Fig. \ref{fig:disc}		\\
921207	&	2083	&	$	0.005	\pm	0.07	$ & $	-1.01	\pm	0.06	$ & $	0.977	\pm	0.06	$ & $	1.23	\pm	0.1	$ & $	-2.8	\pm	0.12	$ & $	1.12	\pm	0.02	$ & $	0.76	\pm	0.04	$	&	pulse 1	\\
921207	&	2083	&	$	0.15	\pm	0.09	$ & $	-0.77	\pm	0.06	$ & $	1.68	\pm	0.19	$ & $	0.98	\pm	0.2	$ & $	-1.5	\pm	0.2	$ & $	1.67	\pm	0.17	$ & $	0.8	\pm	0.11	$	&	a, pulse 2	\\
930112	&	2127	&	$	0.03	\pm	0.15	$ & $	-0.75	\pm	0.16	$ & $	1.86	\pm	0.42	$ & $	1.79	\pm	0.43	$ & $	-3.1	\pm	0.35	$ & $	1.77	\pm	0.14	$ & $	0.71	\pm	0.1	$	&	a	\\
930120	&	2138	&	$	0.14	\pm	0.05	$ & $	-0.43	\pm	0.14	$ & $	1.98	\pm	0.9	$ & $	-0.17	\pm	0.8	$ & $	-1.8	\pm	0.3	$ & $	2.3	\pm	0.3	$ & $	0.14	\pm	0.05	$	&		\\
930201	&	2156	&	$	-0.03	\pm	0.12	$ & $	-0.4	\pm	0.33	$ & $	4.24	\pm	3.27	$ & $	1.75	\pm	0.28	$ & $	-2.7	\pm	0.2	$ & $	3.29	\pm	0.18	$ & $	0.41	\pm	0.07	$	&		\\
930214	&	2193	&	$	-0.25	\pm	0.02	$ & $	-0.78	\pm	0.04	$ & $	12.9	\pm	1.1	$ & $	0.15	\pm	0.07	$ & $	-1.5	\pm	0.1	$ & $	13.1	\pm	0.8	$ & $	0.74	\pm	0.04	$	&		\\
930612	&	2387	&	$	-0.09	\pm	0.05	$ & $	-0.57	\pm	0.06	$ & $	6	\pm	1	$ & $	0.29	\pm	0.07	$ & $	-2.1	\pm	0.2	$ & $	8.8	\pm	0.6	$ & $	0.43	\pm	0.05	$	&		\\
940410	&	2919	&	$	-0.21	\pm	0.2	$ & $	-0.69	\pm	0.1	$ & $	0.85	\pm	0.48	$ & $	-0.16	\pm	0.24	$ & $	-1.13	\pm	0.15	$ & $	0.89	\pm	0.31	$ & $	0.44	\pm	0.13	$	&		\\
940623	&	3042	&	$	-0.05	\pm	0.06	$ & $	-1	\pm	0.2	$ & $	5.3	\pm	0.8	$ & $	0.78	\pm	0.18	$ & $	-2.5	\pm	0.3	$ & $	3.9	\pm	0.2	$ & $	0.34	\pm	0.11	$	&		\\
940708	&	3067	&	$	0	\pm	0.1	$ & $	-0.32	\pm	0.06	$ & $	1.28	\pm	0.46	$ & $	0.01	\pm	0.11	$ & $	-2	\pm	0.25	$ & $	3	\pm	0.4	$ & $	0.07	\pm	0.04	$	&		\\
941023	&	3256	&	$	-0.03	\pm	0.16	$ & $	-0.66	\pm	0.05	$ & $	1.54	\pm	0.43	$ & $	0.05	\pm	0.07	$ & $	-1.28	\pm	0.08	$ & $	2.5	\pm	0.2	$ & $	0.54	\pm	0.09	$	&		\\
941026	&	3257	&	$	-0.18	\pm	0.09	$ & $	-0.4	\pm	0.03	$ & $	2.6	\pm	1.1	$ & $	0.18	\pm	0.1	$ & $	-1.34	\pm	0.08	$ & $	4.38	\pm	0.42	$ & $	0.48	\pm	0.04	$	&		\\
950403	&	3492	&	$				$ & $				$ & $				$ & $				$ & $				$ & $				$ & $				$	&	Complex, Figs. \ref{fig:const}, \ref{fig:T4}\\
950624	&	3648	&	$	-0.04	\pm	0.23	$ & $	-0.79	\pm	0.07	$ & $	1.2	\pm	0.3	$ & $	0.76	\pm	0.93	$ & $	-1.83	\pm	0.6	$ & $	1.41	\pm	0.42	$ & $	0.7	\pm	0.07	$	&	pulse 3	\\
950701	&	3658	&	$	-0.27	\pm	0.08	$ & $	-1.13	\pm	0.15	$ & $	0.81	\pm	0.11	$ & $	0.35	\pm	0.2	$ & $	-4.5	\pm	0.8	$ & $	0.95	\pm	0.08	$ & $	0.72	\pm	0.05	$	&		\\
950818	&	3765	&	$	0.13	\pm	0.05	$ & $	-0.47	\pm	0.05	$ & $	1.3	\pm	0.15	$ & $	0.71	\pm	0.09	$ & $	-2.35	\pm	0.11	$ & $	1.48	\pm	0.06	$ & $	0.02	\pm	0.033	$	&		\\
951016	&	3870	&	$	0.02	\pm	0.16	$ & $	-0.45	\pm	0.07	$ & $	0.52	\pm	0.19	$ & $	0.16	\pm	0.25	$ & $	-1.5	\pm	0.22	$ & $	0.77	\pm	0.2	$ & $	0.22	\pm	0.06	$	&		\\
951102	&	3891	&	$	0.52	\pm	0.76	$ & $	-0.91	\pm	0.5	$ & $	1.04	\pm	0.47	$ & $	2.6	\pm	1.6	$ & $	-2.4	\pm	0.5	$ & $	0.95	\pm	0.17	$ & $	0.41	\pm	0.13	$	&		\\
951213	&	3954	&	$	0.1	\pm	0.11	$ & $	-0.6	\pm	0.1	$ & $	1.13	\pm	0.33	$ & $	0.18	\pm	0.18	$ & $	-1.37	\pm	0.15	$ & $	1.1	\pm	0.2	$ & $	0.01	\pm	0.09	$	&		\\
951228	&	4157	&	$	0.07	\pm	0.06	$ & $	-0.42	\pm	0.03	$ & $	2.5	\pm	0.3	$ & $	0.76	\pm	0.1	$ & $	-1.61	\pm	0.11	$ & $	3.17	\pm	0.14	$ & $	0.87	\pm	0.08	$	&	a, pulse 2	\\
960124	&	4556	&	$	0.07	\pm	0.09	$ & $	-0.83	\pm	0.15	$ & $	1.97	\pm	0.3	$ & $	1.06	\pm	0.2	$ & $	-2.59	\pm	0.3	$ & $	2.01	\pm	0.1	$ & $	0.39	\pm	0.04	$	&		\\
960530	&	5478	&	$	-0.39	\pm	0.07	$ & $	-0.68	\pm	0.08	$ & $	3.4	\pm	1.1	$ & $	0.09	\pm	0.15	$ & $	-1.62	\pm	0.15	$ & $	3	\pm	0.4	$ & $	0.62	\pm	0.04	$	&		\\
960605	&	5486	&	$	-0.1	\pm	0.07	$ & $	-0.82	\pm	0.1	$ & $	1.55	\pm	0.26	$ & $	0.26	\pm	0.18	$ & $	-1.63	\pm	0.23	$ & $	1.66	\pm	0.18	$ & $	0.56	\pm	0.07	$	&		\\
960804	&	5563	&	$	0.04	\pm	0.13	$ & $	-0.83	\pm	0.21	$ & $	0.6	\pm	0.12	$ & $	1.41	\pm	0.25	$ & $	-3.8	\pm	0.6	$ & $	0.7	\pm	0.05	$ & $	0.69	\pm	0.06	$	&	a	\\
960912	&	5601	&	$	-0.17	\pm	0.3	$ & $	-0.93	\pm	0.09	$ & $	2.13	\pm	0.6	$ & $	1.5	\pm	1	$ & $	-1.7	\pm	0.2	$ & $	1.61	\pm	0.29	$ & $	0.76	\pm	0.09	$	&		\\
960924	&	5614	&	$				$ & $				$ & $				$ & $				$ & $				$ & $				$ & $				$	&	Complex, Figs. \ref{fig:const}	\ref{fig:T4}	\\
961102	&	5654	&	$	-0.12	\pm	0.06	$ & $	-0.62	\pm	0.19	$ & $	18.1	\pm	3.5	$ & $	0.78	\pm	0.08	$ & $	-2.6	\pm	0.31	$ & $	18.15	\pm	0.79	$ & $	0.86	\pm	0.08	$	&	a	\\
970111	&	5773	&	$				$ & $				$ & $				$ & $				$ & $				$ & $				$ & $	0.38	\pm	0.01	$	&	Complex; Fig. \ref{fig:R1}		\\
970223	&	6100	&	$	0.09	\pm	0.36	$ & $	-0.46	\pm	0.31	$ & $	1.18	\pm	1.1	$ & $	0.62	\pm	0.4	$ & $	-2.1	\pm	0.3	$ & $	1.3	\pm	0.26	$ & $	-0.02	\pm	0.12	$	&		\\
970420	&	6198	&	$				$ & $				$ & $				$ & $				$ & $				$ & $				$ & $				$	&	Complex; Fig. \ref{fig:R1}		\\
970815	&	6335	&	$	-0.28	\pm	0.02	$ & $	-0.56	\pm	0.03	$ & $	3.51	\pm	0.46	$ & $	-0.29	\pm	0.09	$ & $	-1.38	\pm	0.12	$ & $	3.04	\pm	0.55	$ & $	0.39	\pm	0.01	$	&	subpulse at 3s	\\
970828	&	6350	&	$	0.19	\pm	0.09	$ & $	-0.68	\pm	0.13	$ & $	9.33	\pm	1.14	$ & $	1.6	\pm	0.3	$ & $	-1.93	\pm	0.28	$ & $	9.32	\pm	0.68	$ & $	0.44	\pm	0.05	$	&		\\
970925	&	6397	&	$	-0.21	\pm	0.04	$ & $	-0.55	\pm	0.08	$ & $	6.4	\pm	1.5	$ & $	0.37	\pm	0.11	$ & $	-1.76	\pm	0.11	$ & $	4.18	\pm	0.32	$ & $	0.43	\pm	0.07	$	&	a	\\
971127	&	6504	&	$	-0.11	\pm	0.09	$ & $	-0.67	\pm	0.07	$ & $	3.8	\pm	0.6	$ & $	0.21	\pm	0.1	$ & $	-1.35	\pm	0.08	$ & $	4.7	\pm	0.4	$ & $	0.62	\pm	0.03	$	&		\\
971208	&	6526	&	$				$ & $				$ & $				$ & $				$ & $				$ & $				$ & $				$	&	very weak thermal flux	\\
980306	&	6630	&	$	-0.16	\pm	0.04	$ & $	-1.14	\pm	0.05	$ & $	1.58	\pm	0.09	$ & $	0.31	\pm	0.07	$ & $	-2.7	\pm	0.2	$ & $	1.94	\pm	0.07	$ & $	0.85	\pm	0.06	$	&		\\
980718	&	6930	&	$	-0.3	\pm	0.08	$ & $	-1.3	\pm	0.2	$ & $	2.69	\pm	0.2	$ & $	-0.12	\pm	0.16	$ & $	-2.8	\pm	0.2	$ & $	2.6	\pm	0.2	$ & $	0.6	\pm	0.1	$	&		\\
990102	&	7293	&	$	-0.36	\pm	0.03	$ & $	-0.64	\pm	0.04	$ & $	8.1	\pm	1.3	$ & $	0.4	\pm	0.14	$ & $	-1.78	\pm	0.12	$ & $	5	\pm	1	$ & $	0.52	\pm	0.02	$	&		\\
990102	&	7295	&	$	-0.12	\pm	0.06	$ & $	-0.74	\pm	0.04	$ & $	3.3	\pm	0.5	$ & $	0.83	\pm	0.2	$ & $	-1.46	\pm	0.15	$ & $	2.19	\pm	0.2	$ & $	0.46	\pm	0.07	$	&	excl. subpulse at 10 s	\\
990123	&	7343	&	$	-0.09	\pm	0.04	$ & $	-0.43	\pm	0.15	$ & $	7.5	\pm	2.2	$ & $	0.46	\pm	0.33	$ & $	-1.15	\pm	0.6	$ & $	5.94	\pm	1.9	$ & $	0.25	\pm	0.06	$	&	pulse 1	\\
990316	&	7475	&	$				$ & $	-0.315	\pm	0.04	$ & $				$ & $	0.71	\pm	0.17	$ & $	-2	\pm	0.22	$ & $	10.7	\pm	0.9	$ & $	1.1	\pm	0.22	$	&	b	\\
990424	&	7527	&	$	-0.07	\pm	0.03	$ & $	-0.76	\pm	0.07	$ & $	2.5	\pm	0.2	$ & $	0.08	\pm	0.4	$ & $	-2	\pm	0.6	$ & $	1.8	\pm	0.7	$ & $	0.27	\pm	0.06	$	&		\\
990816	&	7711	&	$	-0.41	\pm	0.06	$ & $	-0.9	\pm	0.07	$ & $	2.6	\pm	0.5	$ & $	-0.01	\pm	0.1	$ & $	-1.74	\pm	0.14	$ & $	2.4	\pm	0.2	$ & $	0.8	\pm	0.03	$	&		\\
991216	&	7906	&	$	0.05	\pm	0.05	$ & $	-0.8	\pm	0.23	$ & $	0.98	\pm	0.16	$ & $	0.94	\pm	0.11	$ & $	-3.2	\pm	0.7	$ & $	1.08	\pm	0.08	$ & $	0.37	\pm	0.04	$	&	pulse 1	\\
  \tablenotetext{a}{Break in ${\cal{R}}$.}
 \tablenotetext{b}{No significant break in the $kT$ evolution}
\enddata
\label{tab:Tt}
\end{deluxetable}

One can argue from a theoretical point of view that the spectrum we identify as a thermal component is formed either by photospheric emission without further dissipative effects (if scattering is the dominating opacity then the emitted emission is rather a Wien spectrum), or by Comptonized thermal emission due to additional dissipation \citep{RM05}.  Furthermore, the actual spectrum is most likely a convolution in time and space of several emitting regions, which, by necessity, broadens and dilutes the spectrum into a diluted and multicolor blackbody (see e.g. Pe'er 2008).  Indeed, it is not statistically possible to distinguish between a pure blackbody, a diluted blackbody, or a Wien spectrum as shown by \citet{Ryde05}.  Therefore,  it is still useful to  fit this component by a Planck function, which in this sense captures such a peaked component, having in mind that this component probably is quasi-blackbody emission.

{Similarly, the single power-law used in the fits, approximates the accompanying broad-band, non-thermal emission. The limited band width prevents a detailed characterization of the non-thermal radiation, in particular its spectral shape.  In a possible scenario, the photons emitted from the photosphere can serve as seed photons for Compton scattering by energetic electrons  produced by dissipation processes in the flow (e.g. \citet{RM05, PMR05}).  If the  dissipation occurs near or below the photosphere, energy exchange via both inverse and direct Compton scattering with the thermal photons may significantly modify the electrons energy distribution, and as a consequence a variety of complicated non-thermal spectra may be obtained \citep{PMR06}. These can not be described by a simple power law over a broad energy range. However,  a single power-law function is a relatively good approximation over  a narrow energy band, such as the BATSE band. See further discussion the validity of this assumption in \S \ref{sec:nonthermal}}. 

\section{Spectral Evolution Analysis of the Prompt Phase Emission}

\subsection{Decomposition into Spectral Components}
\label{sec:decomp}

We show in Figure \ref{fig:1}  the decomposition  of the prompt emission of several BATSE bursts into the two spectral components, namely the thermal and the non-thermal components. These bursts have been chosen to illustrate the spectral dispersion that is observed. The thermal component is prominent and is the cause for the spectral peak, at an observed peak energy $E_{\rm p}  \sim 250$ keV \citep{Kaneko06}. The broad band, non-thermal emission is captured by the power-law component. In this interpretation the spectral evolution is caused by the change in temperature, which can be significant both during the pulse as well as throughout the entire burst.  We find that the  temperature exhibits a characteristic decay behavior, a broken power law in time (see \S 3.2 below). This is similar to the behavior reported by Ryde (2004).  Different combinations of the thermal and the non-thermal components give rise to the variety of observed spectra;  for instance, a variety of subpeak spectral slopes and peak energies. 
Spectral analyses, using the model in equation (1), on bursts observed with other satellite missions (with rather limited band widths) have been reported in for instance,  \citet{mcbreen}, Bosnjak et al. (2006), Falcone et al. (2007), and Bellm et al. (2008).

Our fitting model is an alternative to the Band model, which consists of two smoothly joined power-laws \citep{Band93}. The two-component model gives fits  with reduced $\chi_\nu^2$-values around 1, which is similar to the $\chi_\nu^2$ values found for the  Band model over the BATSE energy band (see Ryde 2004, 2005), however, having a somewhat larger dispersion in values. 
For the sample presented in Table 1, the $\chi_\nu^2$ values vary from 0.85 to 1.15 with typically 110 degrees of freedom per time bin. The averaged reduced $\chi^2$ for the whole sample is $\chi_{\nu}^2 = 1.01$. The lowest $\chi_\nu^2$ values ($<1$) are obtained for time-resolved spectra in which  there is emission at high energies that is not readily captured with a Band function. For these bursts the two-component model gives better fits to the spectra (see also Batellino et al. 2006 and Ryde et al. 2006).   The highest  $\chi_\nu^2$ values are caused by spectra which have large residuals at the lowest and the highest energies. In these cases the single power-law function does not fully capture the broad band emission at the extreme energies (see discussions in \S \ref{sec:nonthermal} for more details). 

Moreover, a majority of spectra fitted with the Band function in the BATSE energy band have $\beta < -2$.  This is, for instance, the case for  BATSE triggers 907 and 1663, shown in Figure 1. Our fits indicate, on the contrary, a rising high energy component.  While these fits have similar $\chi^2$ as the Band function fits (in the BATSE range), their prediction of the flux above 2 MeV are quite different. One reason for the ambiguity of the exact spectral behavior in the BATSE range is the fact that  at high energies the effective area of the detector decreases quickly. The rising spectral slope of course necessitates an additional high energy break.  Broad band observations are needed to clarify this point. This will be further discussed in \S \ref{sec:nonthermal}.

The relative contribution of the thermal emission over the non-thermal emission is different for different bursts.   This is evident from the right-most column in Figure \ref{fig:1},  in which the temporal evolution of the ratio of the thermal to total emission is depicted.  Here the thermal flux is the bolometric integrated blackbody flux and the non-thermal flux is integrated over the energy band $\sim$25-1900 keV. The ratio found, $\sim 30\%- 50\%$, is thus  an upper limit to the true (and unknown) bolometric ratio.  It should be noted that while the relative strength of the thermal component varies in time, no strong recurring trend is revealed. For BATSE trigger 1663 the ratio remains approximately constant, while in other bursts it either increases or decreases.  This is different from the temperature and flux behaviors found below, for which a recurring behaviors are identified. On the other hand, these observations indicate that, over an emission pulse, the thermal and non-thermal fluxes are not independent of each other, since the ratio remains roughly the same as compared  to the flux variation themselves.

\subsection{Temporal  Behavior of the Temperature}
\label{sec:evol}

The temperature of the quasi blackbody component shows a reccurring behavior. During the rise of the pulse, the temperature is typically constant or slowly declining, varying as  a power law in time; the averaged values of the power law index $a_T$ is $<a_T>=-0.07$ with a standard deviation  $\sigma (a_T)=0.19$. After the pulse peak the temperature starts to decay faster, again as a power law;  the average value of $b_T$  is $<b_T>=-0.68$ and $\sigma (b_T)= 0.24$. This is consistent with the findings in the works by Ryde (2004, 2005).  Figure \ref{fig:kT} shows eight  bursts illustrating the variation that is observed. Some pulses have distinct breaks and the broken power-law model gives a univocal representation of the data, illustrated by triggers 1709, 3042, 3765, 4556, 7527 in the figure. Other bursts, however, have a smaller break, that is, the difference $b_T-a_T$ is small.  An unusual behavior is exhibited by BATSE trigger 1883 which does not have a measurable break in its decay at all. The measured values of $a_T$ and $b_T$, as well as the break time, $t_{0,T}$ for the bursts in our sample are given in Table 1.

The panels in the left-most column in Figure \ref{fig:histo} show the distribution of the power law indices $a_T$ and $b_T$ for the sample studied here.  The dispersions are rather large and for  individual bursts these values can be rather different from each other.  On an average, however,  the temperature is constant until the break, at which it starts to decay as power law with an index, which is numerically close to $-2/3$ (see further discussions in  \S 4 and also in  Ryde, 2004).

\subsection{Temporal Variations of the Flux  of Thermal Emission}

For each time interval during a burst, the temperature ($T$)  and the blackbody normalization ($A$) are directly measured from a fit to the  spectral data.  The energy flux ($F_{BB}$) is subsequently calculated.  Although the flux is not independent of the other two quantities, it has a physical meaning of its own and therefore, nonetheless the analysis of the temporal evolution of these three quantities  (using minimization of the $\chi^2$) are made independently of each other. 

During the temperature evolution described above, the energy flux contained in the thermal component typically rises to the peak, which often coincides with the break in the temperature decay, after which it decays off. Figure \ref{fig:flux} shows the thermal fluxes for a number of pulses in our sample. As evident from the figure, the temporal evolution of  thermal fluxes are described by power laws.  The slopes vary from pulse to pulse, and during the rise phase the sample average of the power law index of $<a_{\rm F}> = 0.63$ with a standard deviation $\sigma (a_{\rm F})= 0.66$ (following $F_{\rm BB} \propto t^{-a_{\rm F}}$). During the pulse-decays, the corresponding average value of the indices are $<b_{\rm F}> = - 2.05$ with standard deviation  $\sigma (b_{\rm F})= 0.75$.  The  broad dispersions around the averaged values are prominent in Figure \ref{fig:histo}.

As mentioned above, the break time in the temperature coincides with the peak in thermal flux ($t_{0,F}$)  to within the errors.  This is shown in  the right-most panel in the lower row in  Figure \ref{fig:histo}. The solid line shows a linear relation between the two times. 

\subsection{Normalization of the Blackbody Spectrum} 
\label{34}

In general, the parameter $A(t)$ (see Eq. 1) describes the size of the photosphere from which the blackbody is emitted, as well as the efficiency of the emergent flux (see discussion below). Here, we define a dimensionless parameter ${\cal{R}}$  as ratio of the observed energy flux and the emergent flux (see also Pe'er et al. 2007)
 \begin{equation}
 {\cal{R}}(t) \equiv \left( \frac{F_{\rm BB}(t)}{\sigma T(t)^4}\right)^{1/2}.
 \end{equation}
Therefore, the parameter  ${\cal{R}} (t)$ is proportional to $A(t)^{1/2}$. We now expand previous studies of the photospheric emission by studying the evolution of the parameter ${\cal{R}}(t)$ (instead of $A(t)$), since it contains important information regarding the emission site.

The behavior of ${\cal{R}}$ over four individual pulses is shown in Figure  \ref{fig:Rnew}. The insets show the count light curves over the same time intervals. A common property appears to be the power-law increase of ${\cal{R}}$; ${\cal{R}} (t) \propto t \,^{r}$. Such a behavior is indeed the typical during most bursts,  in particular, during their individual pulses.  The ubiquitousness of this behavior is further shown in Figure  \ref{fig:ratio} , \ref{fig:semi},  \ref{fig:const}. For the bursts presented in Figure \ref{fig:ratio}, the evolution of   ${\cal{R}}$ are modeled by single power laws over the entire pulse duration.  The value of  (the dimensionless) ${\cal{R}}$ often varies over an order of magnitude, with typical values lying around $10^{-19}$ (compare eq. [\ref{eq:R}]). As we will show in \S 4.2,  this corresponds to a physical photospheric radius of
\begin{equation}
 R_{\rm ph} = 2 \times 10^{11} \left( \frac{ {\cal{R}} } {3 \times 10^{-19}} \right) \left( \frac{\Gamma}{300} \right) \left( \frac{d_{\rm L} }{10^{28}{\rm cm}} \right) \left( \frac{1+z}{2} \right)^{-2} {\rm cm}
\end{equation}
The characteristic temporal behavior of ${\cal{R}}$ shown in Figures \ref{fig:Rnew}, \ref{fig:ratio}, \ref{fig:semi}, \ref{fig:const} is very ubiquitous. Since ${\cal{R}}$ is related to the photospheric radius, we find this behavior essential in understanding the properties of the photospheric emission, the emission site and the outflow dynamics.  

In some bursts there is an indication of a break in the power-law increase of ${\cal{R}}$ after a few seconds, after which ${\cal{R}}$ is approximately time independent. Several examples are illustrated in Figure  \ref{fig:semi}. Note that since the flux levels become low at the end of the pulses, where these breaks in ${\cal{R}}$ occur, the error bars on the measured data points are large and consequently they exhibit large dispersion. Therefore, while the break is apparent, the exact behavior of ${\cal{R}}$ after the break is not very clear. Interestingly, we find that this break-time is independent of the break-time in temperature, and can be both leading (e.g. trigger 5654, with coincidence in trigger 2127) and trailing, with the latter being the typical behavior. The power-law indices prior to the break are on average  comparable to those of the single power-law bursts shown above, indicating that the break might have a different origin.

We have also identified a few bursts for which ${\cal{R}}$  is close to being time independent  over the full duration of the pulse. The existence of such bursts  is also evident from the histogram in Figure \ref{fig:histo}, where a few bursts show a power law index $r$ close to 0. Several examples of these bursts are shown in Figure \ref{fig:const}. For BATSE triggers 1625 and 5614, the constant level is reached after  approximately 1 second (a few time bins); see further discussion in \S 4.2.

In the investigation above we have studied individual pulses since these are the main constituents of a light curve and exhibit strong spectral evolution \citep{Band93, RS02}.   However, a large fraction of GRBs have spiky light curves in which individual spikes are too short to enable time resolved analysis. We have nontheless applied our method on a few bursts with highly variable light curves, bearing in mind that the time resolution of the individual time bins do not resolve the light curve variations. The  inferred values of flux and temperature in each time bin are thus averaged values, and as might be expected do not show any clear behaviors. However,  we find that the temporal evolution of ${\cal{R}}(t)$ does show well defined behaviors.
This is illustrated in  Figure \ref{fig:R1} where we show  ${\cal{R}} = {\cal{R}}(t)$ during six bursts of varying complexity, having several spikes which are not individually resolvable. The insets show the count light-curves over the same time intervals.   A remarkable feature that can be seen in the figure, is that for many of these bursts, ${\cal{R}}(t)$ increases more or less independently of the light curve. This can, in particular, be seen for BATSE triggers 1085 and 5773. Again a power-law increase in time is the prevailing behavior. Furthermore, for BATSE triggers 1663 and 6198,  the parameter ${\cal{R}}(t)$  appears to asymptotically settle down to a constant value, during a period in which the light curve is in its most active phase.  These findings  demonstrate, first, that ${\cal{R}}$ is a fundamental quantity and, second, that physical information can indeed be extracted from bursts with highly variable light curves, by studying ${\cal{R}}$. A more extensive study of complex light curves will be presented elsewhere.

The histogram of the distribution of all the power-law  indices $r$ (from fits to ${\cal{R}} (t) \propto t \,^{r}$) is depicted in Figure \ref{fig:histo} (upper, right-most panel).  The distribution does not have a pronounced peak and the averaged  value is  at $<r>=0.51$ while the  standard deviation is $\sigma(r)= 0.25$. 

The power-law fits to the ${\cal{R}}(t)$ data, both over individual pulses as well as complex bursts, often give good fits over the whole observed period (BATSE triggers 2193 and 7711 in Fig. \ref{fig:Rnew} and BATSE triggers 829 and 5773 in Fig. \ref{fig:R1}). However, in some cases the  power law only characterizes the overall (and indeed important) trend (e.g. BATSE trigger 1085 in Fig. \ref{fig:R1}). Variability on top of the power-law fit are apparent (correlated with the peaks in the light curve), but the magnitude of the variability is small relative to the range over which the fit is made. Similar deviations are seen for some of the bursts in Figure \ref{fig:ratio} (e.g. BATSE trigger 6630). More complicated fits (broken power laws etc.) could be made to analyze these secondary features. 

The investigation of the behavior of the parameter ${\cal{R}}$ above can be summarized as follows. Over an individual pulse structure, ${\cal{R}}$ typically increases as a power-law, extending over the  entire pulse duration (exemplified by trigger 2193 in Fig. \ref{fig:Rnew}).  In a few bursts, at the end of the pulse duration,  the parameter ${\cal{R}}$ settles to an approximately constant value (exemplified by trigger 3257 in Fig. \ref{fig:semi}). This break time is independent of the break time in temperature evolution.
For multi-pulse, and complex bursts we find, astonishingly enough, that the power-law can extend over the whole burst independent of individual flux variations (exemplified by BATSE trigger 5773 in Fig. \ref{fig:R1}), while in bursts with well separated pulses  the ${\cal{R}}$ power-law is clearly connected to the individual pulses (exemplified by trigger 2083 in Fig. \ref{fig:ratio}).

\subsection{Comments on the Results}

The fits to the temperature, normalization (${\cal{R}}$), and flux are made independently of each other. Since these parameters are related, consistency between the resulting power laws is expected. We indeed find that the power law indices for the flux, e.g. $b_F$ are close to the  expected value, given the fits to $T$ and ${\cal{R}}$. However, by studying  the power law indices of the decay phase of pulses   in Table 1 it appears, at first glance, that for a few pulses the measured value of $b_F$  deviates somewhat from the  expected value. The reason for this apparent inconsistency is often stochastical fluctuations in the flux data which leads to a measured  break time which is somewhat different from that in the temperature evolution. This in  turn leads to the deviation in the power law temporal slope. More importantly, though, is the late-time break sometimes observed in the ${\cal{R}}$ evolution. Consistency between the parameters should therefore only be expected if one restricts the fitted flux decay phase to earlier times than the break time in ${\cal{R}}$.  However,  it is not a priori obvious from the flux curves that such a restriction should be made; the effect of the break in ${\cal{R}}$ on the flux curve is weak due to the large error bars on the data points. The conclusion is therefore that the fit results of  the time evolution of the quantity $F_{BB}$, which is derived from the basic quantities $A$ and $T$, can give spurious results in some individual bursts.

As illustrated above, typically the flux peak is connected to the break in temperature ($F_{\rm BB} \propto {\cal{R}}^2 T^4$).  In two cases, however, there is no observed break in the temperature (BATSE triggers 1883 and 7475). In these bursts, the flux peak is instead connected to the  the break in ${\cal{R}}$ and is therefore different in origin. In Figure \ref{fig:semi}  the ${\cal{R}}$ evolution in trigger 1883 is also fitted with a broken power law with a break at the flux peak time. Such a fit is indeed consistent with the data.

Finally, an inherent problem in describing data with a power law function is the sensitivity of the value of the index to the chosen zero point in time. In the fits made above the zero point was taken to be at the onset of the pulse. However, in many of the studied bursts the zero point is set by the trigger time of the detector, which does not necessarily exactly coincide with the onset of the pulse. This mainly affects the measured value of the pre-peak power laws, i.e. $a_T$ and $a_F$, which therefore should be treated with some caution. The post peak power laws are not affected by more than typically  {5 \%}, by varying the chosen zero point around the onset of the pulse.

\section{Discussion}

As shown in this paper, we are able to identify the thermal emission component in all of the bursts studied so far,
in the sense that we get acceptable fits to the data.   The characteristic peak in the energy spectra of the prompt emission in GRBs 
is thus interpreted as in many cases stemming from a thermal component. Examples of such spectra are given in, for instance, \citet{Ryde04, Ryde05}.  In other cases, though, the power in the non-thermal components clearly has to peak at a higher energy, which is evident in BATSE triggers 1974 and 7711 (Fig. \ref{fig:1}), and in trigger 7170 (Fig. \ref{fig:7170}).  For the latter cases, the thermal component contributes to the emission of the spectrum below $E\, F_{E}$ peak. The  most important result in our paper is that we  find that the temporal behavior of the thermal component shows clear repetition in all the different bursts. Both the flux of the thermal component and its observed temperature show a broken power-law behavior in time: The temperature is observed to start off at a high level and thereafter it decreases, at first slowly, during the flux rise, and later breaking into a steeper decay, during the decay phase of the pulse.  The break in the temperature occurs in conjunction with the flux peak. The parameter ${\cal{R}}$, reflecting the blackbody normalization,  increases monotonically, typically as a power law in time. Most often, it has the same behavior  over the entire pulse, and sometimes even over the whole burst. 

\subsection{Consequences and Advantages of a Thermal Component Interpretation}
\label{sec:Relevance}

The existence of a photospheric component as part of a braod band spectrum  is indeed expected both in dissipated kinetic outflows \citep{MR00} as well as in dissipative Poynting flux outflows \citep{Drenk, Gi05}.  The optical depth near the base of the flow is enormous, $\tau > 10^{15}$ (for a review, see, e.g., Piran 2005), and therefore  photons emitted by the central engine or by any dissipation mechanism that occurs deep enough in the flow, necessarily thermalize before decoupling from the plasma at the photosphere. In addition, dissipation processes close to  the photosphere (such as internal shocks or magnetic reconnections) can enhance this thermal component by forming a Comptonized spectral peak \citep{RM05}.  The original thermal peak can be upscattered up to a factor of 10 in energy due to the energetic electrons accelerated in the dissipation process.   Further dissipation episodes  expected during the prompt emission  in the optically-thin region above the photosphere produce non-thermal photons.  While the photosphere photons are, in principle, the first to reach the observer, in practice, due to Lorentz contraction the observed time-difference between these photons and non-thermal photons can be shorter than a millisecond and is thus not resolved.  The observer is therefore expected to measure both components simultaneaously. The existence of a thermal spectral component in the prompt emission is thus inevitable in the standard fireball model scenario, and only its relative importance can be debated. 

The interpretation of the spectra as containing a thermal emission component in addition to the non-thermal one, and the finding of the repetition in the temporal behavior of this thermal component have significant advantages over the purely non-thermal interpretation of the spectrum (the ``Band function'').  One obvious result is the ability to apply a consistent physical interpretation to the spectra in cases where the low-energy spectral slope is too steep to account for by the optically-thin synchrotron emission model. Another significant advantage of our interpretation of the data lies in the  assumed physical origin of the thermal emission,
as originating from the photosphere. For instance, the spectral peak has an immediate meaning as an effective (Lorentz-boosted) temperature. In the synchrotron interpretation of the Band model fits, the peak energy is  instead related to the low-energy cut in the distribution of shock-accelerated electrons (characterized by an electron Lorentz factor, $\gamma_{\rm min}$) ; $E_{\rm p} \propto \Gamma \gamma_{\rm min}^2 B'$, where $B'$ is the comoving magnetic field strength (neglecting any dependency on the pitch angle distribution).  There is therefore no a priori reason to expect clustering in $E_{\rm p}$. While non-thermal emission originates from the dissipation of the flow kinetic energy, a mechanism whose details are unknown, much less uncertainty exist in the description of the photosphere. It is thus possible to study the properties of the photosphere, which is the innermost radius from which information can reach us, from study of the thermal emission.

A correlation between the peak energy and flux naturally arises from the properties  of the thermal emission (e.g. \cite{TRM07}). Such correlations are indeed prominent  within pulses and bursts. These are known as the hardness-intensity correlations (or Golenetskii et al. (1983) relations; see also \citet{BR01, KB}). The measured correlations in these works is though between the total flux (thermal + non-thermal) and the spectral peak energy, not only the thermal emission.  However, in our interpretation these correlations originate from the underlying thermal relation, with the  additional non-thermal flux only affecting their appearance. The \citet{amati} correlation represents a corresponding correlation between the flux and $E_{\rm p}$ values of the  {\it time-integrated} spectra for an ensemble of bursts. The relation between temperature and thermal flux might be the underlying cause of this relation as well; see further discussion in \S 5 below. 

In many of the bursts studied here, the spectra are not dominated by the thermal emission, even within the observed energy band (25-1900 keV). We thus suggest, based on the results of this work, that thermal emission can in fact exist in a very large fraction, perhaps even all of the prompt emission spectra of long GRBs.  

\subsection{Radiative Efficiency}

The radiative efficiency of the prompt phase emission is given by the parameter $\eta = E_\gamma/(E_\gamma+ E_{\rm K})$ , where $E_\gamma$ is the radiated, prompt, gamma-ray energy (measured  in the BATSE energy band) 
and $E_{\rm K}$ is the kinetic energy of the fireball right after the prompt phase. The estimation of $E_{\rm K}$ is somewhat dependent  on the invoked afterglow model. However, most estimates point toward very high values of the efficiency, mainly varying from a level of several percent to larger than 90 \% \citep{FW, Z07,N08}. This observational result seems to be in contradiction to the prediction of the internal shock model. In spite of its many successes, a major difficulty in this model is  that, in general, internal shocks are inefficient in tapping the kinetic energy of the flow. Typically, only a few per cent of the fireball energy is converted into gamma-rays \citep{DM98, Kumar99, B00, Sp}. 

Our new interpretation, and in particular the results presented here in Figure 1, may thus help to resolve this issue. According to this interpretation, thermal photons originate from the photosphere and thus they may originate, at least in part, directly from the explosion and not from dissipation of the kinetic energy in the flow. The fact that  the thermal  photons carry a significant fraction ($30\% - 50\%$) of the total observed flux in the BATSE band (see Fig. 1), therefore eases the energy requirements from kinetic dissipative models. This interpretation in supported by numerical simulations  of jet propagation out through the progenitor star by
\citet{L09}.  They find that tangential collimation shocks are important through out  the jet, and that these generate continuous dissipation. This makes the jet internally hot producing a very bright photosphere.  More than half of the total energy in the of the jet can thus be converted into radiation.
  
\subsection{Interpretation of the parameter ${\cal{R}}$}
\label{42}

The measured variations in the observed energy flux represents, to our interpretation,  underlying variations in  the energy input at the central engine\footnote{as opposed to, e.g.,  obscuration.}. The temperature, on the other hand, is given by the energy per particle and also informs us about the thermalization that takes place. While the flux and temperature have rather straight forward  interpretations, we will elude on the interpretation of ${\cal{R}}$ below.

We assume that the flow is advected through a photosphere, at  distance  $R_{\rm ph}$ with  bulk Lorentz factor  $\Gamma>> \Theta_{\rm jet}^{-1}$, where  $\Theta_{\rm jet}$ is the GRB jet opening angle. The observed flux, $F_{BB}$, is given by integrating the intensity over the emitting surface  $F_{\rm BB} = (2\pi/d_{\rm L}^2) \int d\mu\,\mu\, R_{\rm ph}^2 {\cal{D}}^4 (\sigma {T'}^4/\pi)$.  Here $T'$ is the comoving plasma temperature, which is related to the observed temperature via $T = T' \, {\cal{D}}$,
${\cal{D}} \equiv ( \Gamma (1- \beta \mu ))^{-1}$ is the Doppler factor,  $\theta=\rm{arccos}\mu$ is the angle to the line of sight, and 
$\beta = (1-\Gamma^{-2})^{-1/2}$ is the plasma expansion velocity. The ratio ${\cal{R}} \equiv (F_{BB}/\sigma T^4)^{1/2}$ is calculated by integrating over $\theta$:
\begin{equation}
{\cal{R}}  = \xi \, \frac{(1+z)^2}{ \, d_{\rm L}}\,\, \frac{R_{\rm ph}}{\Gamma} , 
\label{eq:R}
\end{equation}
where $z$ is the redshift and $d_{\rm L}$  is the luminosity distance. The coefficient $\xi$ is a numerical factor of the order unity resulting from angluar integration (see \citet{PR07}).    For bursts with known redshift $z$, the parameter ${\cal{R}}$, which is a ratio of observed quantities, can thus be interpreted as an effective transverse size of the emitting region.  

According to this interpretation, a constant ${\cal{R}}$ means that the effective emitting area of the photosphere is time independent.
A few of the bursts in our sample show this property and are thus the easiest to interpret. In Figure \ref{fig:const}  several bursts for which this is the case are presented.  We show in Figure \ref{fig:T4}  the blackbody flux plotted versus its temperature for these bursts. These plots show the hardness-intensity correlations and, as expected, the observed flux is $F_{\rm BB} \propto T^\delta$, with $\delta \eqsim 4$; the fitted values of the power-law indices  are shown in the panels. This is the fundamental property of a blackbody emitter. The existence of this correlation illustrates directly the photospheric interpretation that is put forward here.

We can further demonstrate how to extract information from bursts with complex light curves. Figure \ref{fig:disc} illustrates a specific example, namely GRB921123 (BATSE  trigger 2067). The light curve of this burst consists of several overlapping pulses creating a complex light curve (see Fig. \ref{fig:R1}). The temporal evolution of the temperature and flux  track each other in this particular burst (see also \citet{Crider97}).  The temporal evolution of parameter ${\cal{R}}$ is shown in Figure \ref{fig:R1}. From a statistically point-of-view the data can be fitted by a  single power-law. However, another possible interpretation can be made.  Instead of a single power-law, the  ${\cal{R}}(t)$ data can be represented by two intervals of constant ${\cal{R}}$ values; one value during the interval between $0 -  3$ seconds and another value during the interval between $6 - 15$ seconds. Such an  interpretation is indeed supported by the flux versus temperature plot (right-most panel in Fig. \ref{fig:disc}).  During both these periods the relation between the energy flux and the temperature is $F_{\rm BB} \propto T^q$ with $q {\sim4}$, which is indeed expected when ${\cal{R}}$ is constant. Thus in this interpretation, the effective transverse size of the emitting surface is constant during each time period, but changes between them. This burst illustrates how detailed analysis within the framework presented here can increase our understanding of the behavior of individual bursts.

\subsection{Temporal Evolution of the Thermal Component}

A major finding of this work is the well defined temporal evolution of both the temperature and energy flux of the thermal component.
The physical interpretation of this evolution is not obvious. It could be due to variation in the internal properties of the inner engine that produces the burst outflow (e.g. luminosity, baryon load, mass ejection rate, etc.), which are reflected in the observed temperature and thermal flux.  

 A natural explanation to the late time temporal evolution (after  the temperature break time) was suggested by Pe'er (2008). In this
 work, the properties of the photosphere in relativistically  expanding plasma outflow, characterized by steady Lorentz factor
 $\Gamma \gg 1$ were considered.  It was shown there that the photospheric radius  strongly depends on the angle to the line of sight, $\theta$  (see also  Abramowicz, Novikov \& Paczy\'nski 1991). As a result, thermal photons that decouple from the plasma at high angles  to the line of sight, $\theta \gg \Gamma^{-1}$, can be observed tens of seconds after the thermal photons originating on the line of sight  (for typical parameters characterizing GRB outflows). Moreover, in this work  the standard definition of the photosphere (as a surface in space from which the optical depth $\tau=1$) was extended to consider the probability of thermal photons to undergo their last scattering event before reaching the observer from every point in space in which electrons exist. This enabled calculation  (via probability density functions) of the late time thermal flux and temperature. The results found there, $F_{\rm BB} \propto t^{-2}$ and $T\propto t^{-2/3} - t^{-1/2}$ are remarkably close to  the averaged values that we find for the fitted parameters (see Fig. \ref{fig:histo}).  The broad distribution of the ${\cal{R}}$ parameter found here  may reflect the variation of the temperature evolution.

The excellent agreement found between the theoretical prediction of the  photospheric emission model of Pe'er (2008) and the data presented here, strengthen, to our view, our interpretation that  indeed a thermal emission component exists during the prompt emission phase of GRBs, and that we were able to identify it  correctly. The theoretical model, however, does not predict the early (before  the break in the temperature temporal evolution) temporal  evolution. The nearly flat behavior of the temperature at these times suggests that the emission is dominated by photons emitted on  the line of sight, and that high-latitude emission effects can be  neglected at early times. Thus, according to this interpretation, variation in the light curve during the early times directly reflect variations in  the inner engine activity. This interpretation allowed us (Pe'er et al. 2007) to use data collected during this time to estimate the physical parameters  of the GRB outflow, such as  the size at the base of the flow and the outflow bulk Lorentz  factor.

 We note though, that this interpretation does not naturally explain  the lack of significant break in the temporal evolution of
 ${\cal{R}}$ at early times. Thus, alternative explanations may exist and  further investigations are therefore ongoing.

\subsection{Broad Band Emission of the Non-Thermal Component}
 \label{sec:nonthermal}

While we have focused in this paper on the behavior of the thermal component, a full understanding of the energy release during the prompt phase in GRBs can only be achieved by a complete analysis of both the thermal and non-thermal components. In order to achieve this  one needs a comprehensive analysis of the non-thermal component of time-resolved pulses over a broad energy range. While the observations of the sub-MeV range is very well covered, only a few observations exist at higher energies \citep{atkins,gonz, gonz09, Abdo09}. Moreover, in most cases, these observations are integrated over the entire duration of  pulses and in some cases even over the entire duration of bursts. Since significant spectral evolution occurs during individual pulses, only indirect information regarding the emission processes can be found by studying such duration-integrated spectra. 
We therefore claim that the standard fitting of the GRB prompt emission, namely with the Band function, can only be regarded as a first approximation to the actual spectral form in the observed energy range. In addition as we argue here, for many bursts, the addition of a thermal component is required by the data, thus giving rise to two spectral breaks in the spectrum.

\subsubsection{Indications of multiple spectral breaks in the broad-band spectrum}

The BATSE spectra are often fitted with a Band function. One may thus argue that the broad-band spectral break in the prompt emission ($E \, F_ {\rm E}$)  is at the break seen in the BATSE range, which  on average is at approximately 250 keV \citep{Kaneko06}.  This is however not necessarily always the case. As shown in the spectral catalogue of BATSE bursts by \citet{Kaneko06},  approximately 10\% of all bursts  have, during their entire duration,  time-resolved spectra with a high-energy power-law index $\beta > -2$, that is, increasing power with photon energy ($E \, F_{\rm E} \propto E^{\beta+2}$). Moreover, approximately half of all bursts have periods throughout  their durations over which  $\beta  > -2$. This means that the attributed  peak energy, $E_{\rm p}$, from the sub-MeV data is not  the power peak of the broad-band spectrum, but rather a spectral break. A turn-over (peak) must exist at higher energies for these bursts. Similarly, the relatively low detection rate of  burst emission in the  Fermi LAT energy range,  compared to what is expected from extrapolation of the BATSE $\beta$-values, also indicate additional breaks in the super-MeV spectrum. The broad-band spectra are thus expected to typically have more spectral details than what can be captures by a Band function alone (see also e.g Barat et al. 1989). 

\subsubsection{Broad-band spectral observations (keV-100 MeV) }
 
In a recent work,  \citet{gonz09} have completed the analysis of all 37 {\it CGRO} bursts for which both BATSE and TASC have a clear detection. Out of these bursts 9 have  sufficient temporal resolution, giving a couple of  spectra each (of a few second duration)
during the rising phase of pulses. These are therefore appropriate for direct investigations of the underlying emission processes. 
In the work by \citet{gonz09}  the spectra were fitted  with a Band function and a smoothly broken power law.  Here, we reanalyzed the BATSE data for these nine bursts over the same time intervals. We find that a thermal component can indeed be identified in all cases. An example of our analysis is presented in Figure \ref{fig:7170} which shows the
BATSE energy range ($< 2$ MeV) for GRB 981021 (trigger 7170). This presented spectrum is from an initial time bin, from 1 to 3 s after the trigger. We first present  the Band function fit  (Fig. \ref{fig:7170} left panel). In this fit the high-energy spectral index $\beta =-2.87$ and peak energy are frozen at the values found by \citet{gonz09}.  Note that these values are obtained from their fit to both the BATSE and  TASC data. The low-energy power-law index is  $\alpha= -0.59 \pm 0.01$.  The residuals between the model and the data, presented below the spectrum in the figure, do not show the random distribution that  is expected from  a model that captures the main physical properties of the data. Moreover, the $p$-value of the fit, with $\chi^2({\rm dof})= 1.43  (111)$, is $p=0.003$. This clearly indicates the need for an extra component in the spectrum below its power peak at $\sim 1600$ keV. 

The right-most panel in Figure \ref{fig:7170} shows an alternative spectral fit  to the same spectrum, using a thermal component and a power law\footnote{a smoothly-broken power-law can be used instead, but since $E_p$ is at 1600 keV, the difference is marginal}.  This fit results in  $kT = 160 \pm 5$ keV and a power-law index $s= -1.17 \pm 0.01$, with $\chi ^2 = 1.03$ (109). The distribution of the residuals has now improved, which is also reflected by  $p= 0.38$. Apart from improving the fit, the power law below the non-thermal peak at 1600 keV is now compatible with most non-thermal emission models, e.g. synchrotron emission. In the Band function fit the line-of-death for optically-thin synchrotron emission (slow cooling assumption) was violated ($\alpha > -2/3$). In burst spectra similar to the one presented in Figure \ref{fig:7170}, the non-thermal spectrum has a break at a few MeV, constituting the overall spectral power peak,  while the thermal emission forms a shoulder on the low energy power law.   Such spectra could be underlying the observations in e.g. GRB 900520a (Barat et al.1989) and GRB 080916c (Abdo et al. 2009), which exhibit  spectral breaks at high energies. 

In some of the other bursts analyzed by \citet{gonz09}, the non-thermal component does not have a detected break within the TASC energy range ($E \lesssim 110$ MeV). This is illustrated by the first time bin in GRB920311 (trigger 1473), for which the Band function fit to the BATSE and TASC data gives $\beta = -1.71 \pm 0.08$. The spectral power thus increases with photon energy and exceeds that at the observed spectral break, $E_{\rm p} ^{\rm Band}=  390$ keV. Fitting the BATSE data alone, we find  $kT = 68 \pm 13$ keV and $s=-1.60 \pm 0.05$. The power-law index is consistent with the $\beta$-value found over the TASC range\footnote{Indeed, forcing $s = -1.71$ for our fits over the BATSE range gives only marginally larger $\chi^2$ value and $kT = 83\pm 10$ keV}.  In this burst the spectral break is due to the thermal peak while the non-thermal component is consistent with a single power-law over the energy range 25 keV - 110 MeV. Similarly  in GRB 930506 (BATSE trigger 2329; Kaneko et al. 2008)  there is no evidence for a power peak in the spectrum below 100 MeV.

In many of the spectra taken from these nine bursts, our fits show that  the thermal peak and the non-thermal spectral breaks are relatively close to each other.   Nonetheless,  we stress that these two peaks are not identical. For instance,  in the interval between 1 and 3 seconds of GRB 920311 (1663), we find  $kT = 73$ keV and $E^{\rm non-th.}_{\rm p}= 715$ keV. The model used to fit this spectrum consists of  a Planck function and a smoothly-broken power-law. An indication of such a high-energy break is indeed evident in the fits of this burst in Figure 1. In these bursts the spectral power in the non-thermal component is lower than in the thermal component. As a result it is easy to misinterpret the spectrum as consisting of only a Band function, with a peak at the thermal peak (see Fig. \ref{fig:1}).

As mentioned above, the spectra from the nine bursts studied here are from the initial, rising  phase of the pulses.  As shown by Ryde (2004) the non-thermal power-law typically softens during the duration of a pulse (see also Fig. \ref{fig:1}), and therefore a different broad-band spectral behavior is expected at later times of individual pulses. However, there are no available data for these nine bursts, which sufficiently resolve the decay phase of the pulses.

We point out  that the two spectral breaks identified and  discussed here, one due to $kT$ and the other one due to $E^{non-th.}_{\rm p }$, will by necessity be obscured if the interval studied are time-integrated over significant spectral evolution. This is because both peaks vary in time. Moreover, obliging effects of the derived photon and energy spectra further obscures the apparent spectral shape \citep{fen, bromm}. The $\chi^2$ fitting is namely performed in count space and the photon flux data points, used in the plots,  are estimated based on the best fit model. Different models therefore lead to slightly different derived photon fluxes (note however that the spectral fitting is unaffected by this obliging effect).  This underlines the need for time-resolved and forward folding analysis for a proper understanding of the emission processes during the prompt phase in GRBs.

Even though only limited amount of data are available for time-resolved,  broad-band spectra, and these are mainly for hard bursts (having large  $E_p$ values), we draw an important conclusion. While the dispersion in the energy of the thermal peak is relatively small, the peak energy of the accompanying non-thermal component has a much larger dispersion, ranging from  $\sim 700$ keV to larger than 110 MeV.

\subsubsection{Clues to the origin of the nonthermal component}

We thus conclude that our interpretation, which is an alternative to the Band function, is consistent with the observed data. 
In this work we do not attempt to fully analyze the non-thermal part of the spectrum (this will be presented elsewhere).
However, we briefly review some suggestions on the origin of the non-thermal component. 
In the dissipative outflow model, the non-thermal component is expected to have a non-trivial, broad-band shape: 
The energy dissipation above the photosphere can lead to accelerated particles (emitting synchrotron and/or inverse Compton emission) through  Fermi (stochastic) acceleration, or streaming instabilities \citep{trier}  and/or converter acceleration \citep{deri} in the flow. Plasma oscillations induced through Compton scattering by the burst emission itself, can also give rise to accelerated electrons \citep{trier2}. These lead to a variety of spectral slopes and several possible spectral breaks. 

Furthermore, depending on the optical depth that the photospheric photons experience at the dissipation site the particle distribution will be reprocessed by the photons, thereby altering the emitted spectrum to a varying degree, partly thermalizing the spectrum: Thermal photons could serve as seed photons for IC scattering by energetic electrons which are produced by the dissipation mechanism that dissipates the outflow kinetic energy. In this case, a variety of non-thermal spectra may be produced, of which the exact shape depends on the properties of the unspecified dissipation mechanism \citep{RM05, PMR05, PMR06}.   In particular, the thermal component may or may not dominate the total observed flux, depending on the values of the free model parameters \citep{PMR06}.   Such reprocessing of the non-thermal particle distribution by the photosphere photons introduces a relation between the two components. In addition, since the non-thermal emission is drawn from the kinetic energy of the outflow it should be correlated with the thermal emission, apart from efficiency factors which depend on how and where the dissipations occurs.

 Alternatively, if the outflow is Poynting-flux dominated \citep{Drenk,Gi05}  the dissipation occurs at magnetic reconnection sites throughout the flow. In such a scenario \citet{Gi07} and \citet{Gi08} have shown how the fraction of thermal emission in the spectrum varies with the Lorentz factor of the outflow.
For moderate Lorentz factors a thermal component can claim a substantial fraction ($\sim 30 \%$) while for very large Lorentz factors the fraction is low.   To summarize, in order to understand the largely unknown dissipation processes giving rise to the non-thermal component,  temporally resolved emission pulses, observed over a broad energy range, are required.

\subsection{Further Analyses}

Here we have studied the spectral evolution of a complete sample of long pulses in GRB light curves. We have also studied a sample of 8 bursts with complex light curves and showed in \S \ref{42} and in Figure \ref{fig:R1} that in spite of that the pulses overlap each other, we were able to extract information about the photosphere.  Therefore a more exhaustive study of bursts with  complex light curves is expected to reveal further information. 

On a few occasions, the Swift satellite has provided a broad spectrum (0.2-150 keV) of the prompt phase, with its two instruments BAT and XRT \citep{swift}.  However, to extract physical information on the photosphere  these detections must be sufficiently strong to enable time-resolved spectroscopy in a satisfactory manner. Moreover, a wider spectral range is required. The  Fermi Space Telescope covers an energy range of approximately 10 - 200 GeV with its two instruments (Burst Monitor and the Large Area Telescope) and thus is expected to provide a sufficiently broad energy range (see Band et al., in prep.) to allow this analysis. 

In addition,  the late-time flares, which are observed in the afterglow, can be further studied with this model (see, e.g. \citet{falcone}). 
Once available, using the method described here and in Pe'er et al. (2007), the data will allow to discriminate between competing flaring models, such as slower propagating shells or late-time engine activity. 

The strong photospheric emission identified here can be accompanied by a unique neutrino signal. This takes place if two conditions are met: first, the dissipation of the fireball kinetic energy occurs below or near the photosphere (see Pe'er, M\'esz\'aros \& Rees, 2005, 2006), and second, the dissipation process produces population of energetic protons. The energetic protons then interact with the low energy thermal photons via photo-meson and Bethe-Heitler interactions, as well as proton-proton interactions (see Koers, Pe'er, \& Wijers 2006) to produce high energy neutrino signal. The flux of energetic neutrinos that is produced by these processes depends on the spectral shape of the interacting photons. Thus, a thermal photon spectrum gives rise to a unique neutrino signal (see Wang \& Dai 2008, Murase 2008). The results obtained in these works are  different than former estimates (Waxman \& Bahcall 1997, Dermer \& Atoyan 2003), which considered pure power law spectrum of the interacting photons, neglecting thermal component. A full analysis of the neutrino signal expected during the prompt emission phase has however to consider the full spectral shape of the interacting electrons, hence requires first a full modeling of the prompt emission spectrum (both thermal and non-thermal components).  
  
 \section{Conclusions}

We have identified a thermal component in the prompt spectra of GRBs, with a temperature of a few 100 keV. This component has a characteristic behavior, consistent with that of the flow photosphere. This emission is superimposed on a broad-band, non-thermal emission. The broad band spectrum therefore consists of at least two peaks, or spectral breaks, which are more or less pronounced, one from the photosphere and one from the non-thermal emission.  The power peak  of the of the broad band X-ray and $\gamma$-ray spectrum can either be that of the photosphere or that of the non-thermal component, for strong bursts lying at a few MeV.

We have shown on a sample of 56 long BATSE bursts (25-1900 keV), that it is possible to model instantaneous spectra of the prompt emission with a two-component model, consisting of a Planck function combined with a power law, {approximating the non-thermal component} in the analyzed energy range. Our interpretation implies that the peak in the photon spectrum, at  $\sim 250$~keV, is attributed to the quasi blackbody emission from  the photosphere of the relativistic outflow.  Theoretical arguments that lie in the heart of the GRB fireball model easily imply that a blackbody component is expected. Furthermore, the values of the physical parameters derived from our analysis (e.g. $T$, $F_{\rm BB}$, $R_{\rm ph}$, $\Gamma$; see also Pe'er et al. 2007) are consistent with the predictions of the fireball model \citep{RM00}. In summary, the arguments for this interpretation are the following:

1. {\it Goodness of fits.} From a statistical point-of-view the two-component model fits the data well which is reflected by the reduced $\chi_\nu^2 \simeq 1$ values. In some cases our model gives better fits than the commonly used broken power law (Band et al. 1993) model  (see \S 3). In other cases the Band model gives a better $\chi_\nu^2$-value, indicating the need for a more complex model (other than a single power law) to fit the non-thermal component.

2. {\it Radiative Efficiency}. While the non-thermal photons originate from dissipation of the kinetic energy, e.g., by internal shock waves, or magnetic reconnection, thermal photons can  originate directly from the progenitor, i.e., photons that are released in the explosion and thermalize before decoupling at the photosphere. As we have shown in Figure 1, thermal photons carry $30\% - 50\%$ of the total flux. Thus, inclusion of the thermal photons in the total energy budget of the prompt emission can contribute to explaining the high efficiency reported, which is difficult to account for by theoretical models  of kinetic energy dissipation.

3. {\it Recurring behavior.} The behavior of the thermal emission component is similar for most bursts, showing a
particular and recurring behavior. For the 49 pulses we studied we found that both the temperature, $T(t)$, and the flux of the thermal component, $F_{\rm BB}(t)$, exhibit a well defined, and most importantly repetitive behaviour: a broken power law in time. 
During the pre-peak phase the temperature is approximately constant while the energy flux rises. During the post peak-phase, both the flux and the temperature decay as power-laws in time;  $F_{\rm BB} \propto t^{b_F}$ , and $T \propto t^{b_T}$. While the power-law indices do have a broad distribution as shown in Figure \ref{fig:histo} and Table 1, the sample-averaged value of the temperature indices is  $<b_T> = -0.68$ and the average of the flux indices is $<b_F> = -2.05$. The repetitive behaviour found, together with the excellent agreement between the results found here and the theoretical predictions of Pe'er (2008), strongly support, to our view, the  interpretation of the prompt emission spectra as containing a thermal component. 

4. {\it Blackbody normalization.} The normalization of the thermal emission (parameterized by 
${\cal{R}}(t) \equiv (F_{BB}/\sigma  T^4)^{1/2}$) shows a distinct, also recurring,  temporal behavior which is largely independent of the temperature evolution. We found that the parameter ${\cal{R}}(t)$ typically  increases in size as a power law in time over the whole pulse and sometimes even over the entire duration of bursts with complex light curves.  We have thoroughly discussed (\S \ref{34}, \S \ref{42}, Figs. \ref{fig:Rnew}, \ref{fig:ratio}, \ref{fig:semi}, \ref{fig:const}, \ref{fig:R1}) the evolution of  ${\cal{R}}(t)$ and  have shown that this parameter, which is directly derived from  observed quantities, is directly related to the photospheric radius. Hence, it carries with it a significant information on the physics of the prompt emission. In principle, the photosphere is the innermost radius from which any photon can reach the observer; therefore, studying emission from the photosphere holds the key to our understanding of the prompt emission and the properties of the site from which it emanates.

5. {\it Spectral shapes.} Our interpretation of the prompt emission alleviates many of the problems of purely non-thermal interpretations, such as the optically-thin  synchrotron interpretation. In particular, hard sub-peak spectra that challenge the optically-thin synchrotron interpretation of the prompt emission spectra are naturally obtained in our interpretation.   

6. {\it Peak energy.} Identification of the sub-MeV spectral peak energy with the thermal component $\sim kT$, implies that it is independent of the local magnetic field strength and particle acceleration mechanisms, in contrast to the non-thermal (synchrotron) interpretation. This facilitates explanations of observed peak energy clustering and spectral evolution. In the analysis carried out in \S \ref{sec:nonhermal}, we show that the break in the non-thermal component, accompanying the thermal emission, is observed to have a larger dispersion as may be expected by theory.

7. {\it Spectral correlations.} Thermal emission gives a natural correlation between temperature and flux. This correlation should be underlying the hardness-intensity correlation (HIC or Golenetskii relation) observed within bursts, as well as the Amati-like relations which are found for ensembles of bursts and therefore relate them to each other. We note that for several of the studied bursts in this work the observed flux varies as the fourth power of the observed temperature, which is expected from a blackbody emitter. 
The implication for the Amati-like relations, however, should be taken with some care: First, these relations are derived for time-integrated spectra, which is different from the time-resolved spectra considered here. And second, we expect Componization of the thermal component to alterate the position of the peak (Pe'er et al., 2006). Quantitative studies of this alteration require knowledge of the distribution of the non-thermal electrons produced by the dissipation process close to the  photosphere. This, in turn, can be modeled by study of the non-thermal part of the spectra. A detailed study is on-going.

We conclude that the characteristic behaviors that exist for the temperature and the ${\cal{R}}$ evolutions are  the defining properties of the photosphere, and must therefore hold the key to our understanding of the prompt emission and the properties of the site from which it emanates.

\begin{acknowledgments}
We wish to thank Milan Battelino, Peter M\'esz\'aros, Sinead McGlynn, Martin J. Rees, and Ralph Wijers for
many useful discussions and contributions to the analysis. AP is supported by the Riccardo Giacconi
fellowship award of the Space Telescope Science Institute. AP would like to thank the Physics Department at the Royal Institute of Technology in Stockholm  for their hospitality during his visit when part of this project was performed.  Financial support from the Swedish National Space Board is acknowledged with thanks.

\newpage

\begin{figure}[!ht]
\includegraphics[width=1\textwidth]{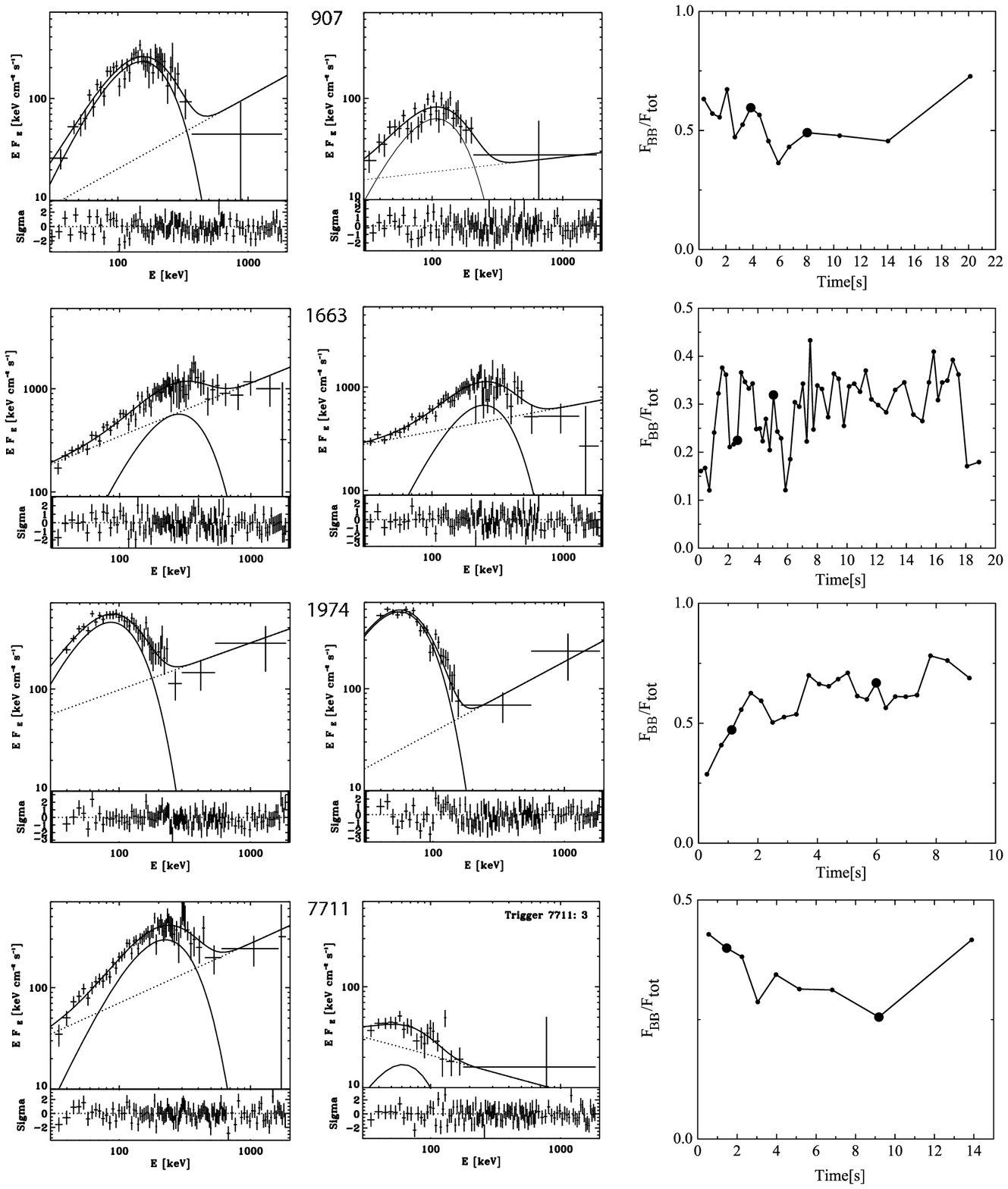}
\caption{Examples of time resolved $E\,F_{E}$ spectra, observed by BATSE, fitted with the two-component model; solid line is the Planck function and the dashed line is the non-thermal (power law) model. The small panels beneath each spectrum show the residuals in units of the standard deviation, $\sigma$. The spectra are rebinned to have a signal-to-noise ratio of unity to make the plots clearer. The full resolution is however kept for the residuals panels.  For every burst, which is identified by its BATSE trigger number, two time-bins are shown in the first two columns.  BATSE trigger 1663 has a complex light curve, while the other three bursts consist of a distinct pulse (see Table 1). These cases illustrate the variation of spectral shapes among bursts, as well as within bursts. 
In the right-most column, the temporal variation of the ratio between the thermal and the total fluxes (in the $\sim$ 25-1900 keV band) is shown for the burst whose spectra are presented  to the left (time bins marked by large dots). The thermal flux is calculated by integrating the Planck function, while the non-thermal flux is calculated by integrating over  the energy band $\sim$20-2000 keV. 
\label{fig:1}}
\end{figure}

\begin{figure}[!ht]
\includegraphics[width=1.0\textwidth]{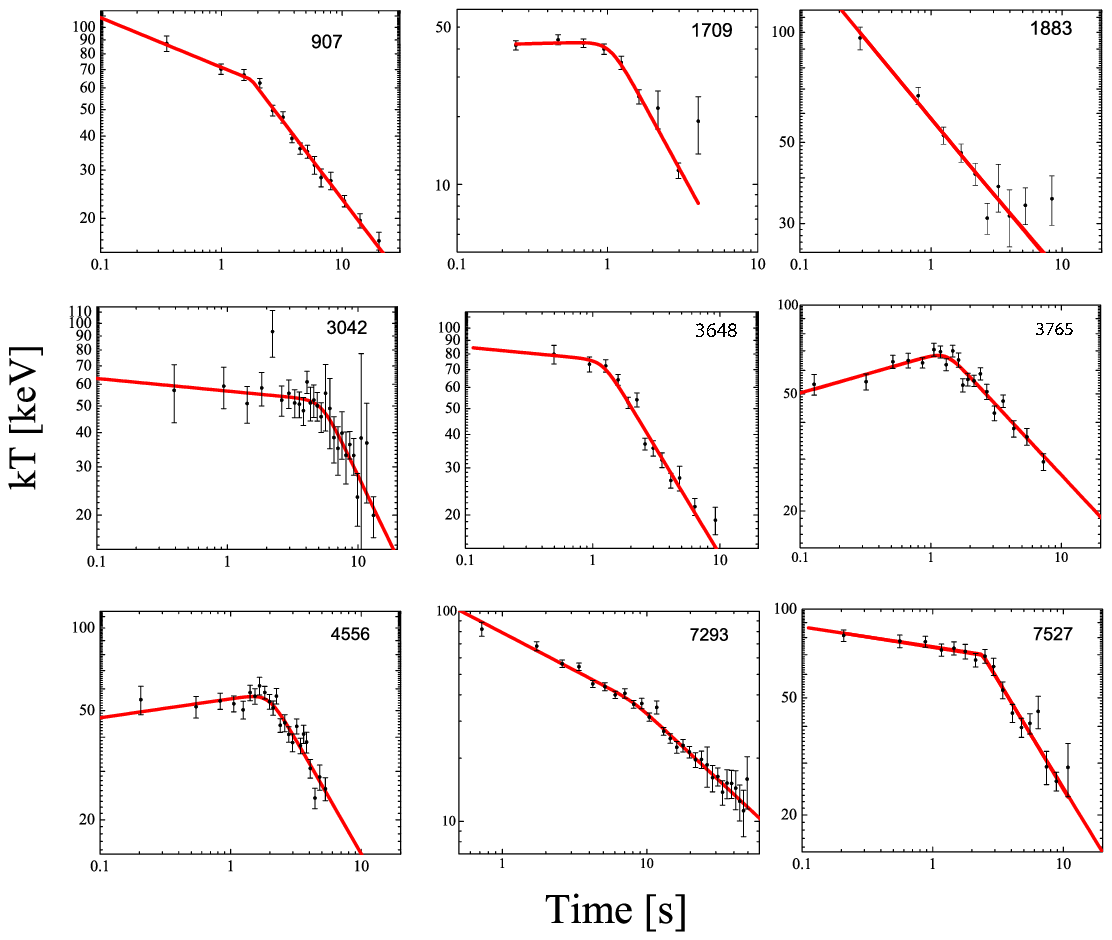}
\caption{Examples of temperature evolution from individual pulses in different bursts. A distinct break is here seen in all cases but trigger 1883 which is best fitted by a single power-law. The slopes vary somewhat from pulse to pulse, but the general behavior is universal.\label{fig:kT}}
\end{figure}

\begin{figure}[!ht]
\includegraphics[width=1.0\textwidth]{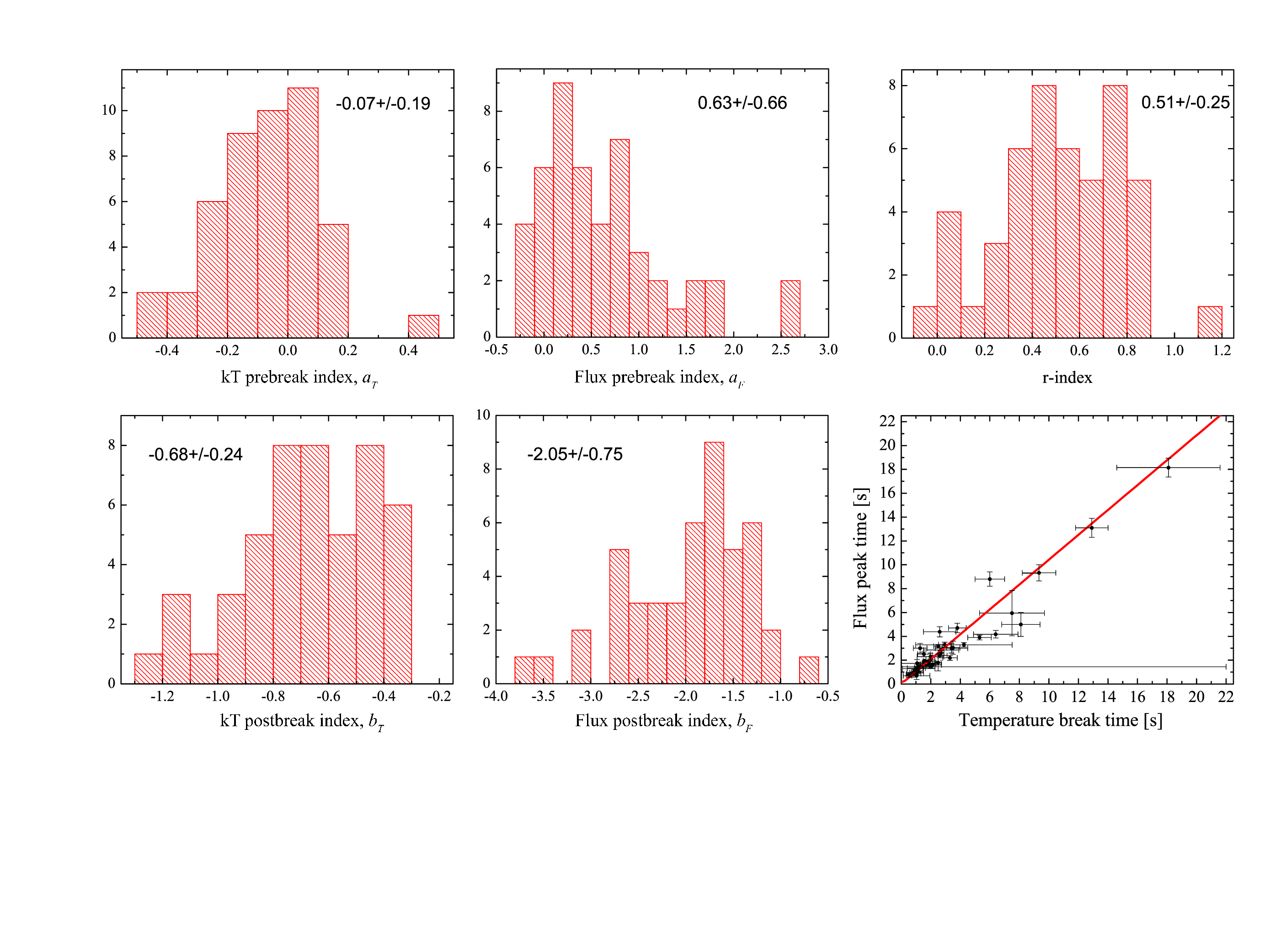}
\caption{Parameter distributions of the analyzed bursts. The panels in the left-most column depicts the histogram of the power-law indices, before ($a_T$) and after ($b_T$) the break in the temperature curve. The column in the middle correspondingly depicts  the indices before ($a_{\rm F}$) and after ($b_{\rm F}$) the flux peak. The upper right-most panel depicts the histogram of the power law index of the ${\cal{R}}$ evolution. Finally, the lower right-most panel shows the correlation between the temperature break time and the flux peak time. The red line shows a linear relation. See the text for further details. \label{fig:histo}}
\end{figure}

\begin{figure}[!ht]
\includegraphics[width=1.0\textwidth]{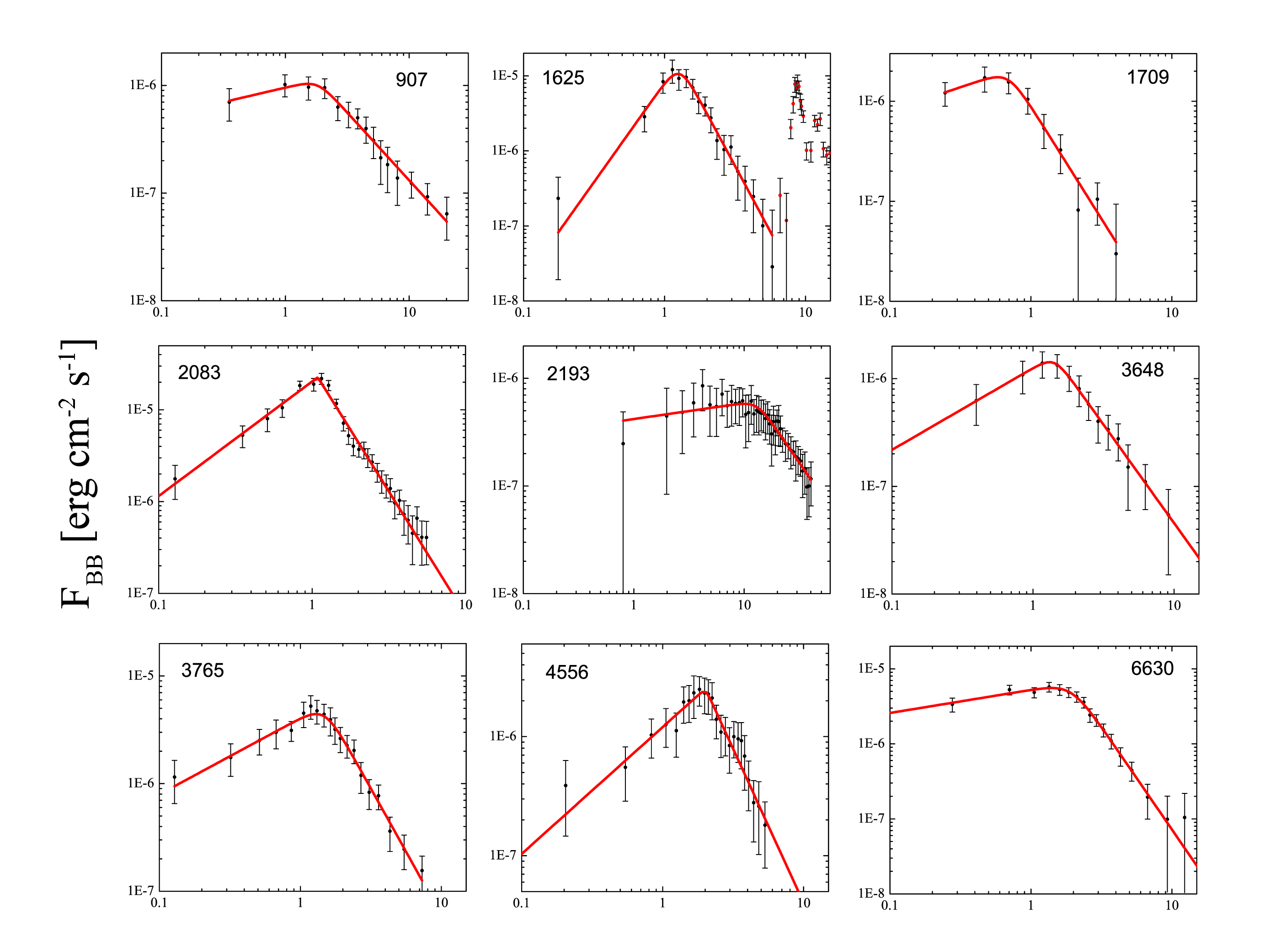}
\caption{Examples of blackbody flux evolution from individual pulses in different bursts. A distinct peak is seen in all cases. Both  the rise and decay phases are describes by power laws.  Note that the actual rise index is very dependent on the zero point of the abscissa. In many cases the trigger is such that data is only available after part of the flux rise has occurred. The red dots refer to additional pulses that are excluded in the presented fits. 
\label{fig:flux}}
\end{figure}

\begin{figure}[!ht]
\includegraphics[width=1.0\textwidth]{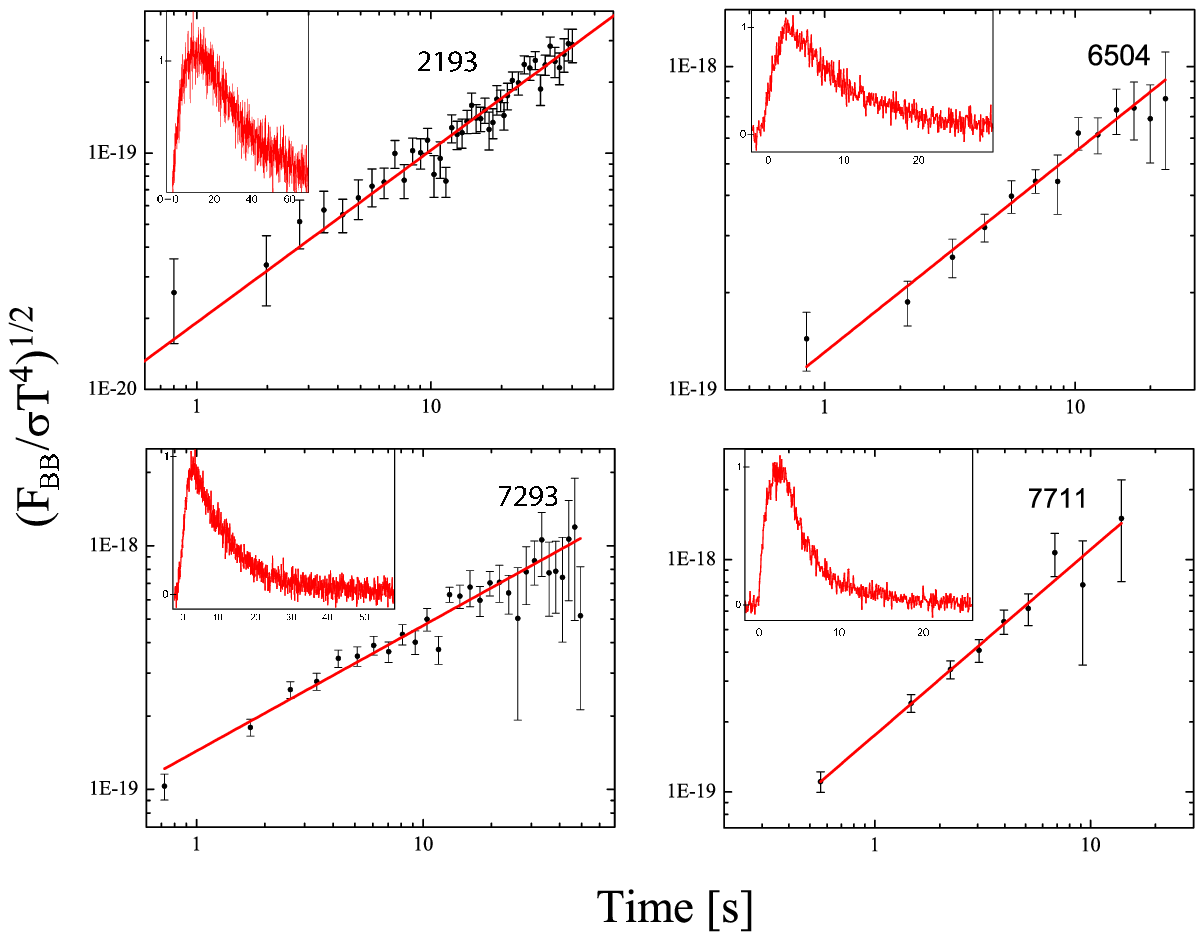}
\caption{Evolution of the parameter ${\cal{R}} = (F_{BB}/\sigma T^4)^{1/2}$ describing the ratio between observed, thermal flux and the emergent flux.  The corresponding count light curves are shown as insets (arbitrary units). A remarkable power law is exhibited, much independent of the rise and decay of the pulse. \label{fig:Rnew}}
\end{figure}

\begin{figure}[!ht]
\includegraphics[width=1.0\textwidth]{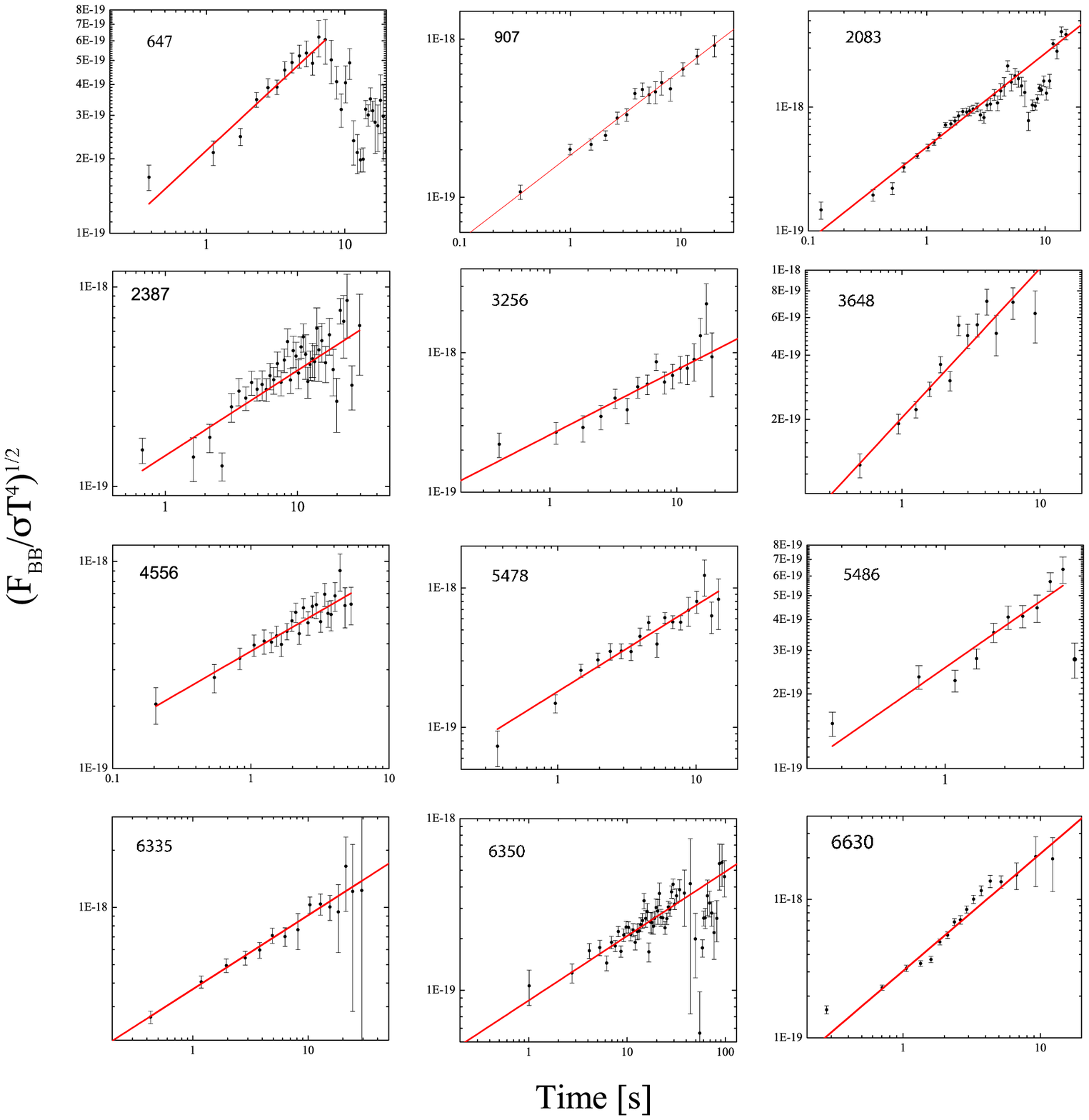}
\caption{More examples of the evolution of the parameter ${\cal{R}} = (F_{BB}/\sigma T^4)^{1/2}$ describing the ratio between observed flux and the emergent flux for individual pulses.  A universal behavior is seen in all bursts. They exhibit a power-law increase over the whole pulse. \label{fig:ratio}}
\end{figure}

\begin{figure}[!ht]
\includegraphics[width=1.0\textwidth]{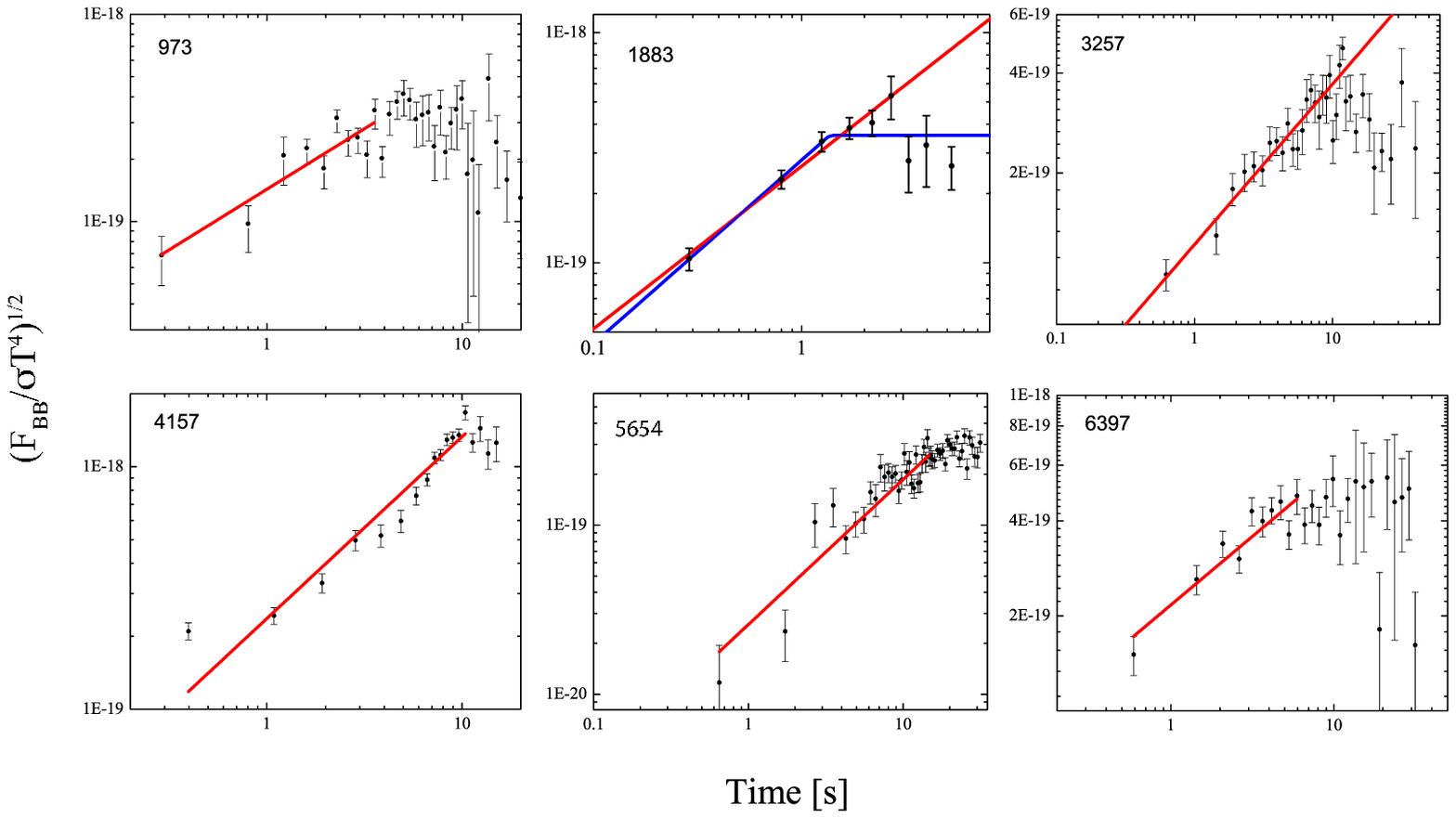}
\caption{Examples of the evolution of the parameter ${\cal{R}}$ for burst with an apparent change in the power law index $r$.  These bursts initially  exhibit a power-law increase, similar to the bursts shown in Figs. \ref{fig:Rnew} and \ref{fig:ratio}. In these bursts, after a few seconds the increase ceases and ${\cal{R}}$ becomes close to constant.  \label{fig:semi}}
\end{figure}

\begin{figure}[!ht]
\includegraphics[width=1.0\textwidth]{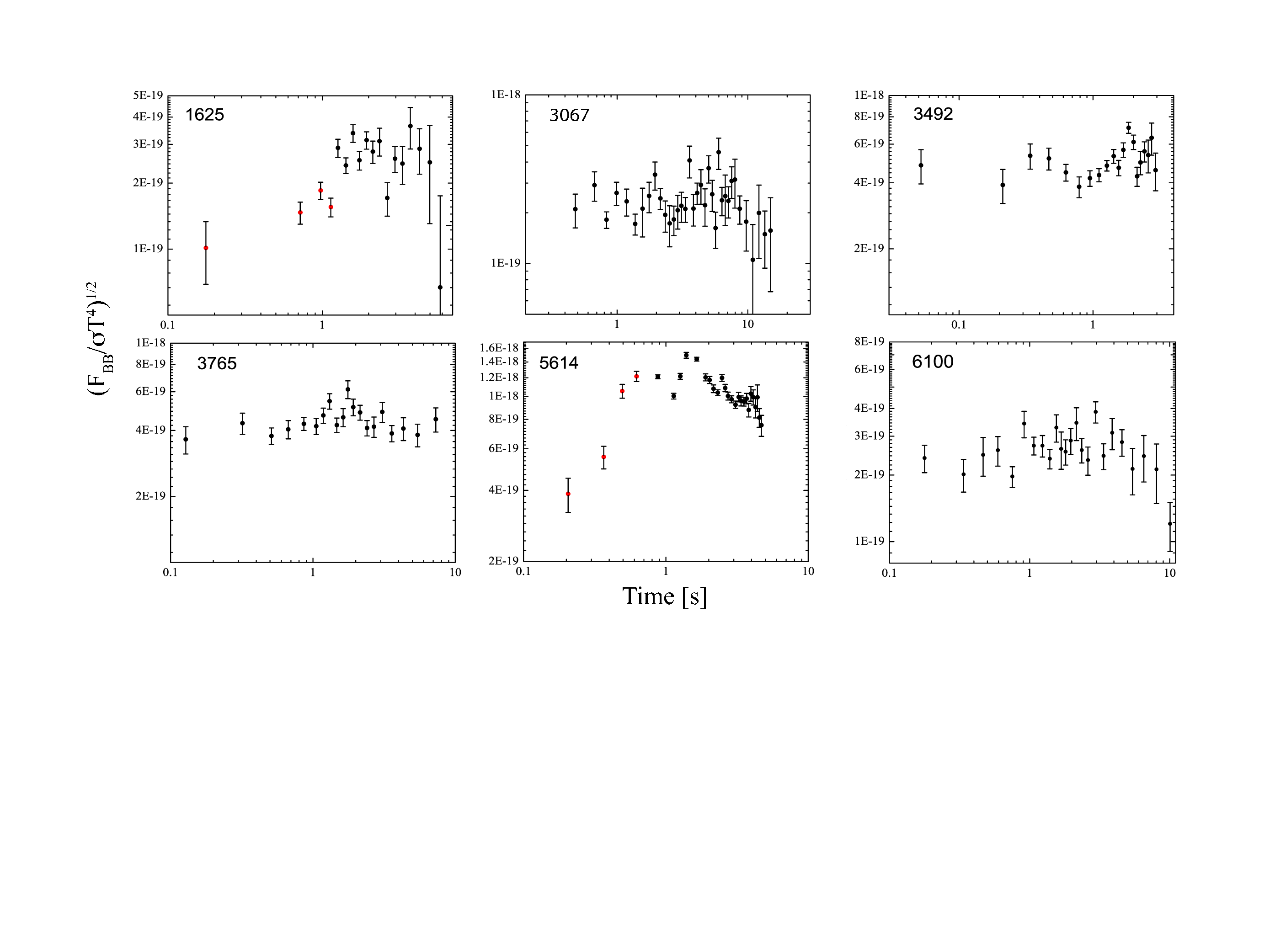}
\caption{Examples of the evolution of the parameter ${\cal{R}}$   
in bursts for which $r$ is close to zero. These bursts are indeed consistent of having ${\cal{R}}$ being constant throughout the analyzed time period. The data points marked by red dots in triggers 1625 and 5614 are probably part of  a short initial increase. The light curves of triggers  3492 and 5614 reveal a complex structure (see Fig. 6 in \citet{RS02}). 
 \label{fig:const}}
\end{figure}

\begin{figure}[!ht]
\includegraphics[width=1.0\textwidth]{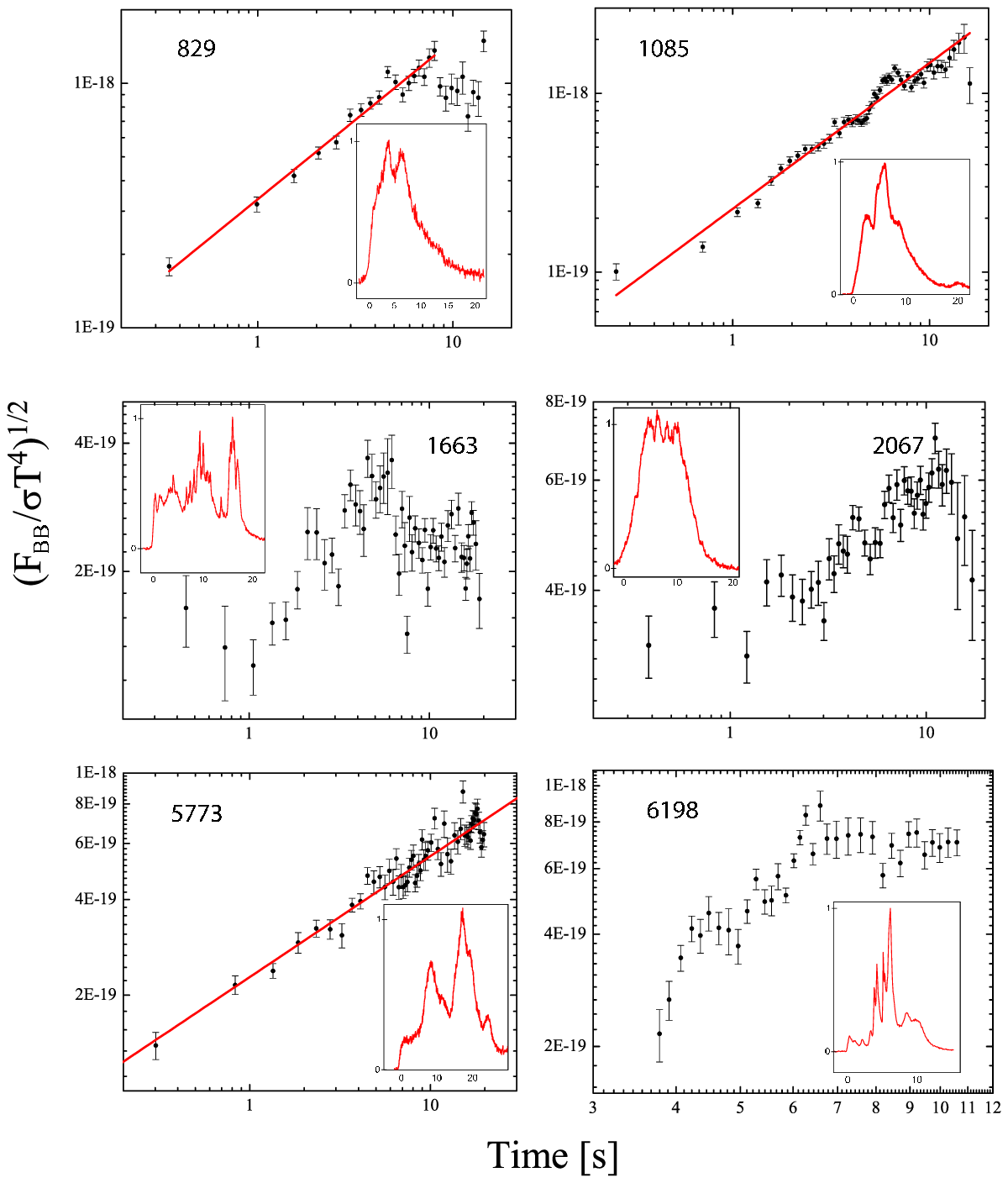}
\caption{Evolution of the parameter ${\cal{R}} = (F_{BB}/\sigma T^4)^{1/2}$ for bursts with complex and heavily overlapping light curves.  The corresponding count light curves are shown as insets (arbitrary units). A remarkable power law is exhibited in many of the bursts, much independent of the complexity of the light curve.  \label{fig:R1}}
\end{figure}

\begin{figure}[!ht]
\includegraphics[width=1.0\textwidth]{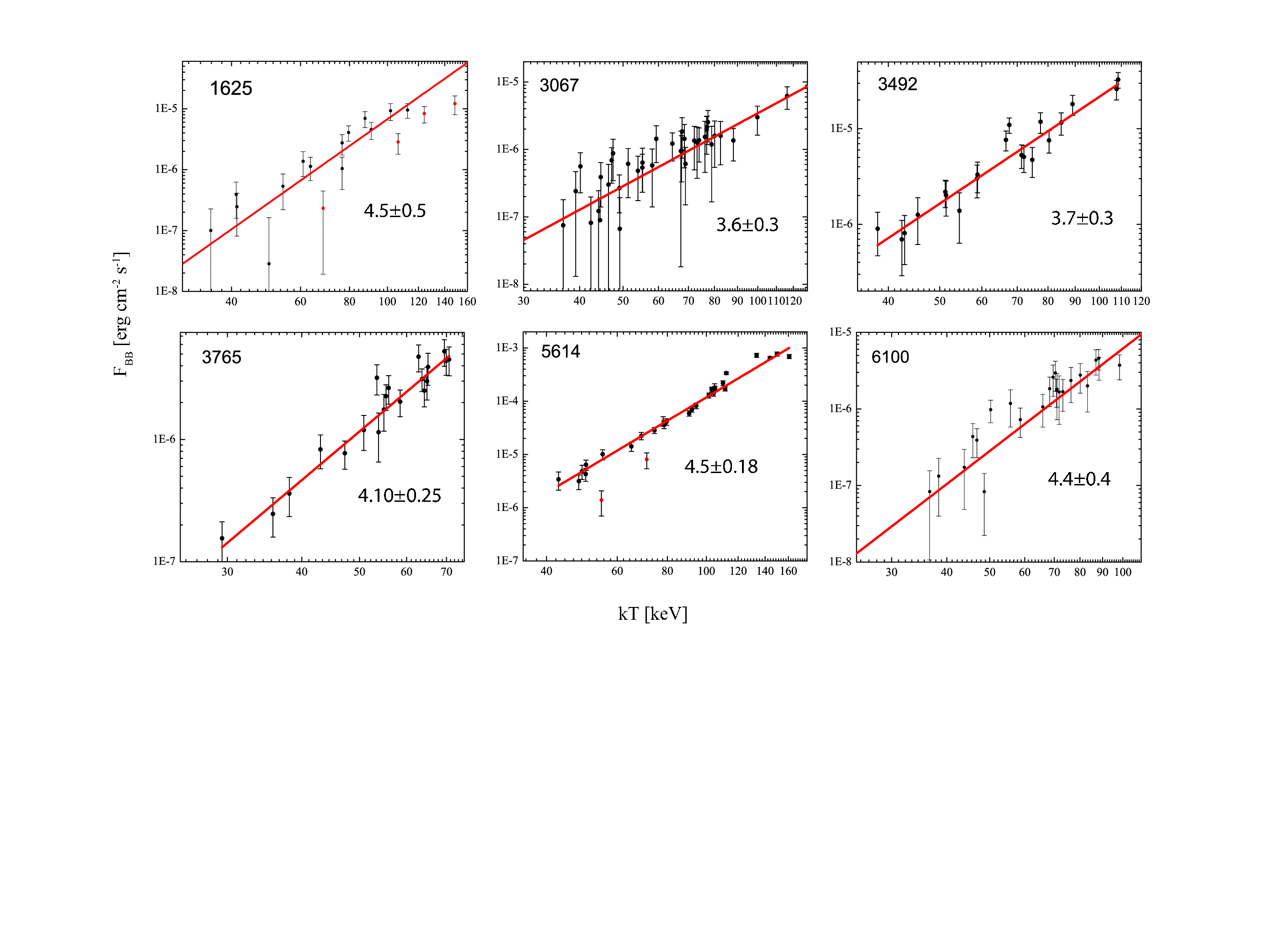}
\caption{Plots of the observed blackbody flux, $F_{\rm BB}$, as a function of  the temperature, $kT$, for the bursts presented in Figure \ref{fig:const}. Such correlations are also known as the hardness-intensity correlations \citep{BR01}. The energy flux is found to increase to the fourth power of the temperature, and is therefore proportional to the emergent flux, $\sigma T^4$. The values of the individual power law fits are shown in the panels. The existence of such bursts illustrate directly the photospheric interpretation of the spectral peak. \label{fig:T4}}
\end{figure}

\begin{figure}[!ht]
\includegraphics[width=1.0\textwidth]{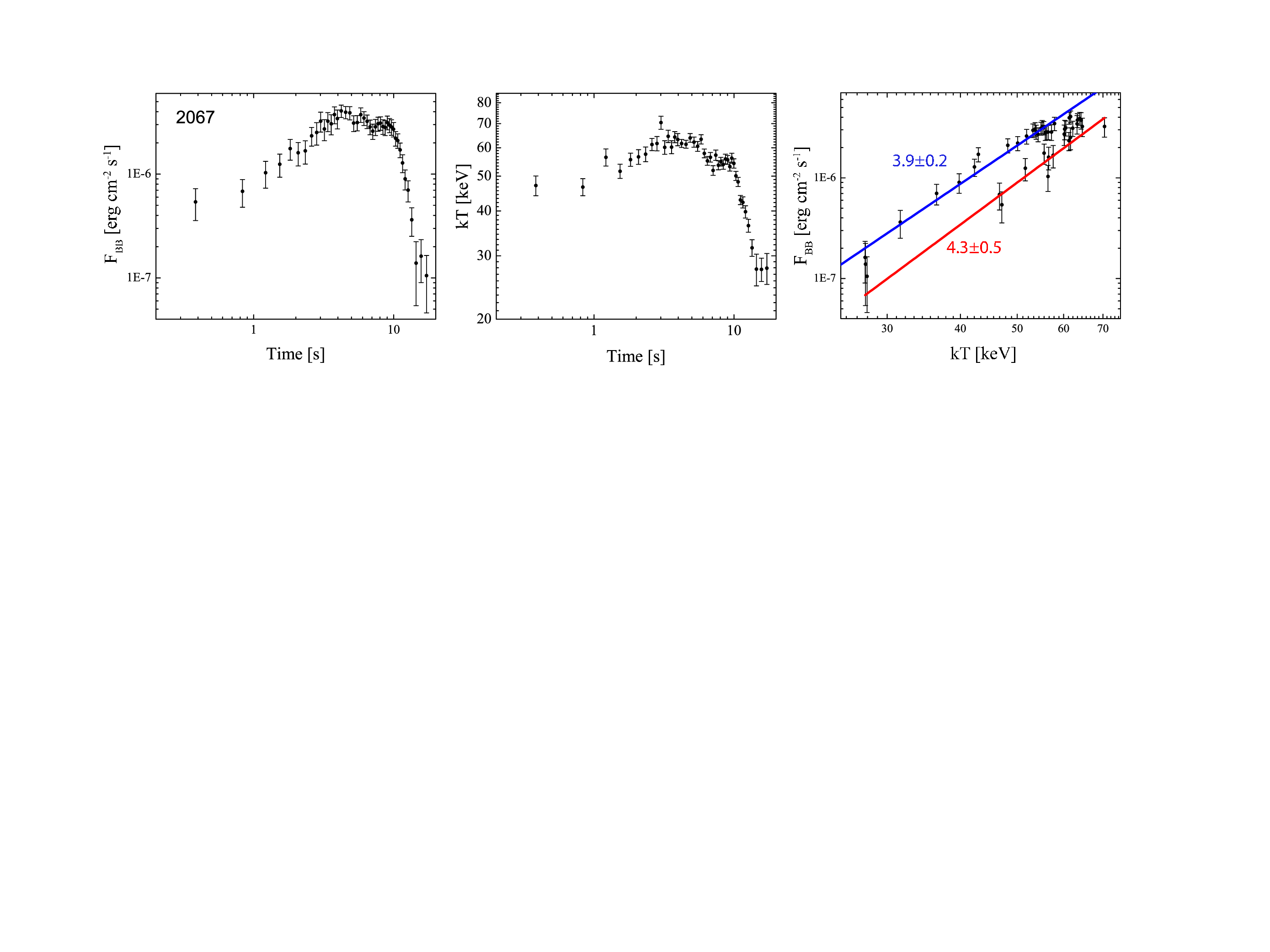}
\caption{Spectral evolution of the blackbody emission in BATSE trigger 2067. Its complex light curve is shown in Figure \ref{fig:R1}. (a) Thermal flux light curve. (b) Temperature evolution. (c) Flux versus temperature.  Further discussion is given in the text. \label{fig:disc}}
\end{figure}

\begin{figure}[!ht]
\includegraphics[width=1.0\textwidth]{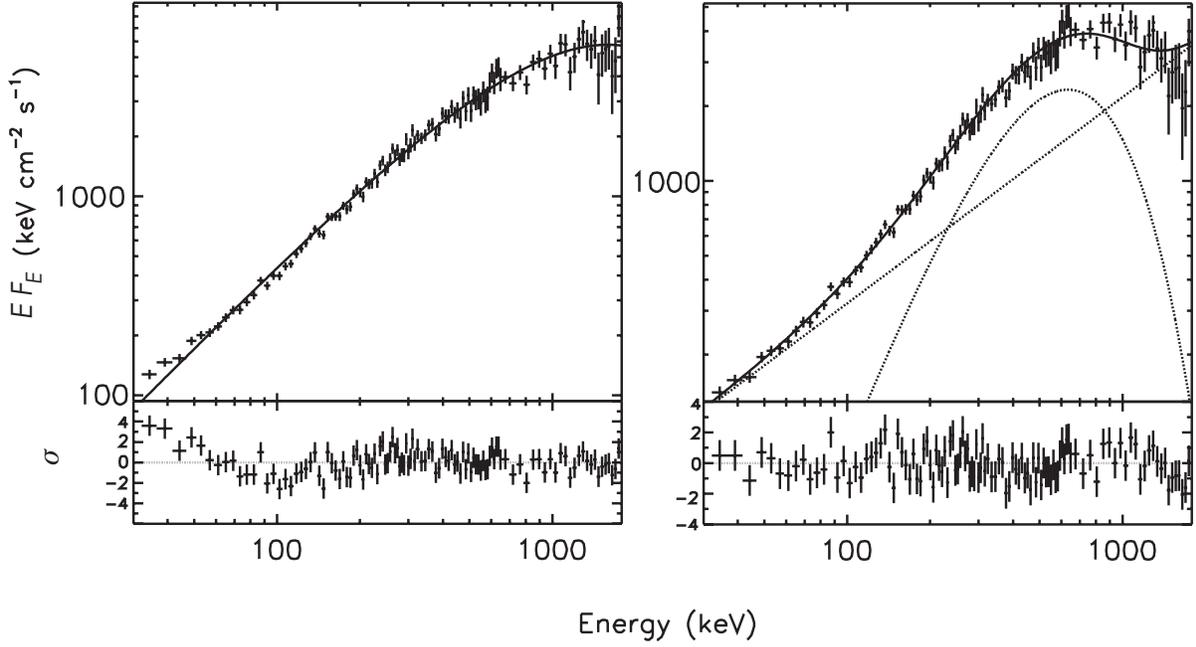}
\caption{$E \, F_{\rm E}$  spectrum from an initial time bin (1-3 s) in GRB981021 (BATSE trigger 7071). 
The panel to the left shows the spectrum fitted to the Band function with $\beta$ set to -2.87 and $E_{\rm p}=1591$ keV found from the TASC fit  \citep{gonz09}. The residuals clearly indicate the need for an extra component in the spectrum ($\chi^2= 1.43$ (111)). Moreover, the fitted value of $\alpha= -0.59 \pm 0.01$ violates the synchrotron line-of-death ($\alpha > -2/3$). 
The panel to the right shows the spectral fit with $kT = 160 \pm 5$ keV and power law index $s= -1.17 \pm 0.01$, giving $\chi ^2 = 1.03$ (109).  The fit is improved and the non-thermal component now has $\alpha < -2/3$, which is consistent with synchrotron emission. The change in flux scale is due to the obliging effect of the deduced photon spectra (see the text for further discussion). \label{fig:7170}}
\end{figure}

\end{acknowledgments}


\begin{thebibliography}{}

\bibitem[Abdo et al. (2009)]{Abdo09} Collaborations, T.~F.~L., et al.\ 2009, Science, 323, 1688 
\bibitem[Abramowicz et al.(1991)]{A91} Abramowicz, M.~A.,  Novikov, I.~D., \& Paczynski, B.\ 1991, \apj, 369, 175 
\bibitem[Amati et al. (2002)]{amati} Amati et al. 2002, A\&A, 390, 81
\bibitem[Atkins et al. (2000)]{atkins} Atkins, R., et al. 2000, ApJ, 533, L119
\bibitem[Band et al. (1993)]{Band93} Band, D., et al. 1993, ApJ,  413, 281 
\bibitem[Barat et al. (1998)]{bar98}Barat, C., Lestrade, J. P., Dezalay, J-P, Hurley, K., Sunyaev, R., Terekhov, O., \& Kuznetsov, A. 1998, in AIP Conf. Proc. 428, Gamma-ray Bursts, 4th Huntsville Symposium,  ed. C. A. Meegan, R. D. Preece \& T. M. Koshut (New York: AIP), 278
\bibitem[Baring \& Braby (2004)]{BaringB}Baring, M.G., \& Braby,  M.L. 2004, ApJ, 613, 460
\bibitem[Barraud et al. (2003)]{Barraud03} Barraud, C., et al.\ 2003, \aap, 400, 1021
\bibitem[Beloborodov(2000)]{B00} Beloborodov, A.~M.\ 2000,  \apjl, 539, L25 
\bibitem [Blinnikov et al. (1999)]{BKP99} Blinnikov, S.I., Kozyreva, A.V., \& Panchenko, I.E. 1999, Astron. Rep., 43, 739
\bibitem[Borgonovo \& Ryde (2001)]{BR01} Borgonovo, L., \& Ryde, F. 2001, \apj, 548, 770
\bibitem[Bosnjak et al.(2006)]{bosnjak} Bosnjak, Z., Celotti, 
A., \& Ghirlanda, G.\ 2006, \mnras, 370, L33 
\bibitem[Bromm \& Schaefer(1999)]{bromm} Bromm, V., \& Schaefer, B.~E.\ 1999, \apj, 520, 661 
\bibitem[Butler et al.(2007)]{KB} Butler, N.~R., Kocevski, D., Bloom, J.~S., \& Curtis, J.~L.\ 2007, \apj, 671, 656 
\bibitem [Cohen et al. (1997)]{Cohen97} Cohen, E., et al. 1997, \apj, 488, 330
\bibitem[Crider et al. (1997)]{Crider97}  Crider, A., et al. 1997, ApJ, 479, L39 
\bibitem [Crider et al. (1998)]{CLP98} Crider, A., Liang, E.P., \& Preece, R.D. 1998, AIPC, 28, 359
\bibitem[Daigne  \& Mochkovitch(1998)]{DM98} Daigne, F., \& Mochkovitch, R.\ 1998, \mnras, 296, 275
\bibitem[Daigne \& Mochkovitch (2002)]{DM02} Daigne, F. \& Mochkovitch, R. 2002, MNRAS, 336, 1271
\bibitem[Derishev et al.(2008)]{deri} Derishev, E.~V.,  Aharonian, F.~A., Kocharovsky, V.~V.,  \& Kocharovsky, V.~V.\ 2008, International Journal of Modern Physics D, 17, 1839 
\bibitem[Dermer  \& Atoyan(2003)]{DA03} Dermer, C.~D., \& Atoyan, A.\ 2003, Physical Review Letters, 91, 071102 
\bibitem[Drenkhahn(2002)]{Drenk} Drenkhahn, G.\ 2002, \aap, 387, 714 
\bibitem[Falcone et al.(2007)]{falcone} Falcone, A.~D., et al.\  2007, \apj, 671, 1921
\bibitem[Fenimore et al.(1983)]{fen} Fenimore, E.~E., Klebesadel, R.~W.,  \& Laros, J.~G.\ 1983, Advances in Space Research, 3, 207 
\bibitem[Fishman et al. (1989)]{Fish89}Fishman, G.J. et al. 1989, in Proc. {\it Gamma Ray Observatory} Science Workshop, ed. W.N. Johnson (Greenbelt: GSFC), 2
\bibitem[Freedman  \& Waxman(2001)]{FW} Freedman, D.~L., \& Waxman, E.\ 2001, \apj, 547, 922 
\bibitem[Frederiksen et al.(2004)]{trier} Frederiksen, J.~T.,  Hededal, C.~B., Haugb{\o}lle, T., \& Nordlund, {\AA}.\ 2004, \apjl, 608, L13 
\bibitem[Frederiksen(2008)]{trier2} Frederiksen, J.~T.\ 2008, \apjl, 680, L5 
\bibitem[Frontera et al. (2000)]{Frontera00} Frontera, F., Amati, L., Costa, E. et~al.\  2000, ApJSS, 127, 59
\bibitem[Gehrels et al. (2004)]{swift}Gehrels, N., Chincarini, G., Giommi, P., et al. 2004, ApJ, 611, 1005
\bibitem[Giannios(2008)]{Gi08} Giannios, D.\ 2008, \aap, 480, 305 
\bibitem[Giannios  \& Spruit(2007)]{Gi07} Giannios, D., \& Spruit, H.~C.\ 2007, \aap, 469, 1 
\bibitem[Giannios  \& Spruit(2005)]{Gi05} Giannios, D., \& Spruit, H.~C.\ 2005, \aap, 
\bibitem[Golenetskii et al. (1983)]{gol}Golenetskii, S. V., Mazets, E. P., Aptekar, R. L., \& Ilyinskii, V. N.
1983, Nature, 306, 451
\bibitem[Gonz{\'a}lez et al.(2003)]{gonz} Gonz{\'a}lez, M.~M., Dingus, B.~L., Kaneko, Y., Preece, R.~D., Dermer, C.~D., \& Briggs, M.~S.\ 2003, Nature, 424, 749
\bibitem[Gonz{\'a}lez et al.(2009)]{gonz09} Gonz{\'a}lez,  M.~M., Carrillo-Barrag{\'a}n, M., Dingus, B.~L., Kaneko, Y., Preece, R.~D., 
\& Briggs, M.~S.\ 2009, \apj, 696, 2155 
\bibitem[Giannios  \& Spitkovsky(2009)]{gi09} Giannios, D., \& Spitkovsky, A.\ 2009, arXiv:0905.1970 
\bibitem[Ghirlanda et al. (2003)]{ghirlanda}Ghirlanda, G.,  Celotti, A., \& Ghisellini, G. 2003, A\&A, 406, 879
\bibitem[Ghirlanda et al. (2004)]{ghir_rel}Ghirlanda, G., Ghisellini, G., \& Lazzati, D. 2004, ApJ, 616, 331
\bibitem[Ghirlanda et al. (2007)]{ghirlanda07} Ghirlanda, G., Bosnjak, Z., Ghisellini, G., Tavecchio, F.,  \& Firmani, C.\ 2007, MNRAS, 379, 73 
\bibitem[Ghisellini et al. (2000)]{g00}Ghisellini, G., Celotti, A., Lazzati, D. 2000, MNRAS 313, L1
\bibitem[Ghisellini et al. (2000)]{ghis00} Ghisellini, G., Lazzati, D., Celotti, A., \& Rees, M. J. 2000, MNRAS, 316, L45
\bibitem[Horv{\'a}th et  al.(2006)]{horvath} Horv{\'a}th, I., Bal{\'a}zs, L.~G., Bagoly, Z., Ryde, F., \& M{\'e}sz{\'a}ros, A.\ 2006, \aap, 447, 23 
\bibitem[Hurley et al.(1994)]{hurley}{Hurley~K. et al.} 1994,{Nature},{372},{652}
\bibitem[Kaneko et al. (2006)]{Kaneko06} Kaneko, Y., et. al. 2006, \apjs, 166, 298
\bibitem[Kaneko et al.(2008)]{Kaneko08} Kaneko, Y.,  Gonz{\'a}lez, M.~M., Preece, R.~D., Dingus, B.~L.,  \& Briggs, M.~S.\ 2008, \apj, 677, 1168 
\bibitem[Kobayashi et al.(1997)]{K97} Kobayashi, S., Piran, T., \& Sari, R.\ 1997, \apj, 490, 92 
\bibitem[Koers et al.(2006)]{KPW06} Koers, H.~B.~J., Pe'er, A., \& Wijers, R.~A.~M.~J.\ 2006, arXiv:hep-ph/0611219
\bibitem[Kouveliotou et al. (1993)]{kouv}Kouveliotou, C. et al. 1993, ApJ, 413, L101
\bibitem[Kumar (1999)]{Kumar99}Kumar, P. 1999, ApJ, 523, L113
\bibitem[Lazzati et al.(2009)]{L09} Lazzati, D., Morsony, B.~J., \& Begelman, M.\ 2009, arXiv:0904.2779 
\bibitem[Lazzati et al.(1999)]{L99} Lazzati, D.,  Ghisellini, G., \& Celotti, A.\ 1999, \mnras, 309, L13 
\bibitem[McBreen et  al.(2006)]{mcbreen} McBreen, S., Hanlon, L., McGlynn, S., McBreen, B., Foley, S., Preece, R., von Kienlin, A., \& Williams, O.~R.\ 2006, \aap, 455, 433 
\bibitem[M\'esz\'aros (2002)]{mesrev} M\'eszar\'os, P. 2002, ARA\&A, 40, 137
\bibitem [M\'esz\'aros (2006)]{Mes06}  M\'esz\'aros, P. 2006, Rep. Prog. Phys., 69, 2259
\bibitem[M\'esz\'aros, \& Rees (2000a)]{MR00}  M\'eszar\'os, P.,\&  Rees, M. J. 2000,  ApJ, 530, 292
\bibitem[M{\'e}sz{\'a}ros \& Rees(2000b)]{541} M{\'e}sz{\'a}ros, P., \& Rees, M.~J.\ 2000, \apjl, 541, L5 
\bibitem[M\'esz\'aros et al. (2002)]{MRRZ02}  M\'eszar\'os, P., Ramirez-Ruiz, E., Rees, M. J., \& Zhang, B. 2002, ApJ, 578, 812
\bibitem[Murase(2008)]{M08} Murase, K.\ 2008, Physical Review D 78, 101302
\bibitem[Nysewander et al.(2008)]{N08} Nysewander, M.,  Fruchter, A.~S., \& Pe'er, A.\ 2008, arXiv:0806.3607
\bibitem[Paciesas et al. (1999)]{batse}Paciesas, W.S., et al., ApJS, 122, 465 
\bibitem[Pe'er(2008)]{P08} Pe'er, A.\ 2008, \apj, 682, 463 
\bibitem [Pe'er et al. (2005)]{PMR05}  Pe'er, A.,  M\'esz\'aros, P., \& Rees, M.J. 2005, \apj, 635, 476
 \bibitem [Pe'er et al. (2006)]{PMR06} Pe'er, A.,  M\'esz\'aros, P., \& Rees, M.J. 2006, \apj, 642, 995
 \bibitem[Pe'er et al. (2007)]{PR07} Pe'er, A., Ryde, F.,  Wijers, R.~A.~M.~J., M{\'e}sz{\'a}ros, P.,\& Rees, M.~J.\ 2007, \apjl, 664, L1
\bibitem[Pe'er \& Waxman (2004)]{PW}Pe'er, A., \& Waxman, E. 2004, ApJ, 613, 448
\bibitem [Preece et al. (1998a)]{Preece98a} Preece, R.D., et al. 1998, \apj, 496, 849
\bibitem[Preece et al. (1998b)]{Preece98b} Preece, R.D., Briggs, M.S., Mallozzi, R.S., Pendleton, G.N., Paciesas, W.S., \& Band, D.L. 1998, ApJ, 506, L23
\bibitem[Preece et al. (2000)]{Preece00}  Preece, R. D., Briggs, M. S.,Mallozzi, R. S., Pendleton, G. N., Paciesas, W. S., \& Band, D. L. 2000, ApJSS, 126, 19
\bibitem [Preece et al. (2002)]{Preece02} Preece, R.D., et al. 2002, \apj, 581, 1248
\bibitem[Rees \& M\'esz\'aros (2000)]{RM00}  Rees, M. J. \& M\'eszar\'os, P.  2000, ApJ,
\bibitem[Rees \& M\'esz\'aros (2005)]{RM05}  Rees, M. J. \& M\'eszar\'os, P.  2005, ApJ, 628, 847
\bibitem[Ryde (2004)]{Ryde04}   Ryde, F. 2004, ApJ, 614, 827
\bibitem[Ryde (2005)]{Ryde05}   Ryde, F. 2005, \apjl, 625, L95
\bibitem[Ryde (2008)]{Ryde08}  Ryde, F. 2008, Phil. Trans. R. Soc. A, 366, 4405
\bibitem[Ryde \& Battelino (2005)]{RB}Ryde, F., \& Battelino, M. 2005, Il Nuevo Cimento, 28C:3, 335
\bibitem[Ryde et al.(2006)]{ryde06} Ryde, F., Bj{\"o}rnsson,  C.-I., Kaneko, Y., M{\'e}sz{\'a}ros, P., Preece, R.,  \& Battelino, M.\ 2006,  \apj, 652, 1400 
\bibitem[Ryde \& Petrosian(2002)]{RP02} Ryde, F., \& Petrosian, V.\ 2002, \apj, 578, 290
\bibitem[Ryde \& Svensson(2002)]{RS02} Ryde, F., \& Svensson, R.\ 2002, \apj, 566, 210 
\bibitem[Stern \& Poutanen(2004)]{SP04} Stern, B.~E., \& Poutanen, J.\ 2004, \mnras, 352, L35 
\bibitem[Schaefer et al. (1998)]{Schaefer98} Schaefer, B.E., et al. 1998, \apj, 492, 696 
\bibitem[Spada et al.(2000)]{Sp} Spada, M., Panaitescu, A., \& M{\'e}sz{\'a}ros, P.\ 2000, \apj, 537, 824
\bibitem[Spitkovsky(2008)]{spit08} Spitkovsky, A.\ 2008,  \apjl, 682, L5 
\bibitem[Strohmayer et al. (1998)]{stro98}Strohmayer, T. E., Fenimore, E. E., Murakami, T., \& Yoshida, A. 1998, \apj, 500, 873
\bibitem[Tavani (1996)]{Tavani96}  Tavani, M. 1996, ApJ, 466, 768
\bibitem[Thompson(2006)]{2006ApJ...651..333T} Thompson, C.\ 2006, \apj, 651, 333 
\bibitem[Thompson et al. (2007)]{TRM07} Thompson, C.,  M{\'e}sz{\'a}ros, P., \& Rees, M.~J.\ 2007, \apj, 666, 1012 
\bibitem[Wang  \& Dai(2008)]{WD08} Wang, X.-Y., \& Dai, Z.-G.\ 2008, arXiv:0807.0290 
\bibitem[Wang et al.(2009)]{Wang09} Wang, X., et al.\ 2009,	arXiv:0903.2086v1
\bibitem[Waxman  \& Bahcall(1997)]{WB97} Waxman, E., \& Bahcall, J.\ 1997, Physical Review Letters, 78, 2292
\bibitem[Zhang et al.(2007)]{Z07} Zhang, B., et al.\ 2007,  \apj, 655, 989 
\bibitem[Zhang \&  M\'esz\'aros (2002)]{ZM}Zhang, B. \&  M\'esz\'aros, P. 2002, ApJ, 581, 1236
\end{thebibliography}
\end{document}